\documentclass[preprint,12pt]{elsarticle}
\usepackage[margin=1in]{geometry} 
\usepackage{amsmath,amsthm,amssymb}
\usepackage{graphicx}
\usepackage{color}
\usepackage[dvipsnames]{xcolor}
\usepackage{subcaption}
\usepackage[export]{adjustbox}
\usepackage{algorithm}
\usepackage{lmodern}
\usepackage[noend]{algpseudocode}
\usepackage{tikz}
\usepackage[toc,page]{appendix}
\usetikzlibrary{bayesnet}

\graphicspath{ {./pic/} }

\DeclareMathOperator*{\argmin}{\text{argmin}}

\begin{document}
	
\title{ MFPC-Net: Multi-fidelity Physics-Constrained Neural Process}

\author[a]{Yating Wang}
\author[a]{Guang Lin\corref{cor1}}
\cortext[cor1]{Corresponding author}
\ead {guanglin@purdue.edu}
\address[a]{Department of Mathematics, Purdue University, West Lafayette, IN 47907, USA}


\begin{abstract}
Recently, there are numerous work on developing surrogate models under the idea of deep learning. Many existing approaches use high fidelity input and solution labels for training. However, it is usually difficult to acquire sufficient high fidelity data in practice. In this work, we propose a network which can utilize computational cheap low-fidelity data together with limited high-fidelity data to train surrogate models, where the multi-fidelity data are generated from multiple underlying models. The network takes a context set as input (physical observation points, low fidelity solution at observed points) and output (high fidelity solution at observed points) pairs. It uses the neural process to learn a distribution over functions conditioned on context sets and provide the mean and standard deviation at target sets. Moreover, the proposed framework also takes into account the available physical laws that govern the data and imposes them as constraints in the loss function. The multi-fidelity physical constraint network (MFPC-Net) (1) takes datasets obtained from multiple models at the same time in the training, (2) takes advantage of available physical information, (3) learns a stochastic process which can encode prior beliefs about the correlation between two fidelity with a few observations, and (4) produces predictions with uncertainty. The ability of representing a class of functions is ensured by the property of neural process and is achieved by the global latent variables in the neural network. Physical constraints are added to the loss using Lagrange multipliers. An algorithm to optimize the loss function is proposed to effectively train the parameters in the network on an ad hoc basis. Once trained, one can obtain fast evaluations at the entire domain of interest given a few observation points from a new low-and high-fidelity model pair. Particularly, one can further identify the unknown parameters such as permeability fields in elliptic PDEs with a simple modification of the network. Several numerical examples for both forward and inverse problems are presented to demonstrate the performance of the proposed method.

\end{abstract}


\maketitle

\section{Introduction}
\label{sec:intro}

Many applications in science and engineering may encounter parameterized computational models, such as optimization of complex systems with uncertainty or inverse modeling. A common challenge lies in obtaining sufficient high-fidelity data, which requires expensive measurements or simulations for a large number of parameter instances. Many approaches have been proposed to develop surrogate models, such as global and local model reduction techniques \cite{ab05, egw10, GMsFEM13, ee03, abdul_yun, fish2008mathematical, matache2002two, NLMC}. The idea is to construct computational cheap reduced order models, however, the simulation data obtained may be low-fidelity and these approaches may face the challenge in generalization.

On the other hand, numerous works have been proposed to learn surrogate models given some data. In the case when only the data are available without knowing the underlying physics, one can treat it as a pure data-driven supervised learning task using methods like Gaussian process (GP) \cite{rasmussen2003gaussian, damianou2013deep}. To obtain reliable results, one usually needs sufficient high-fidelity training data. However, if the random coefficients in the equation possess uncertainties and are high dimensional, or if the underlying map is nonlinear, GP may be inefficient and computationally expensive due to the curse of dimensionality. In the past few years, deep neural networks (DNNs) \cite{schmidhuber2015deep} have attracted increasing interest due to the universal function approximation property and their ability to model high-dimensional input-output relationships. Numerous studies are proposed to solve partial differential equations and have shown great performance in various applications. For example, learning the evolution operator of the PDE from data \cite{wu2020data}, approximating important physical quantities of the PDE \cite{Ying_paraPDE}, solving heterogeneous elliptic problems on varied domains\cite{winovich2019convpde}, designing efficient algorithms to handle multiscale multiphase flow problems \cite{wang2020efficient} and using the idea of multiscale model reductions for learning \cite{wang2018NLMCDNN,wang2020reduced,cheung2020deep}.

When the physical equations/laws are also available, it is important to incorporate these information in the design of surrogate deep neural networks to speed up learning. Physics informed neural networks (PINN) \cite{PINN1,PINN2} were proposed to realize the idea and have been successfully applied to solve PDE problems subject to the law of physics that govern the data. On the other hand, for the case without data labels in the training, there have also been developments on physics-constrained surrogates for stochastic PDEs \cite{PCDL_nz} just using the information on physical relationships. These approaches either require a large amount of high-fidelity data to train, or needs to know the full high-fidelity information of coefficients in the model, such as random coefficients in the equation throughout the entire domain. This is usually hard to realize in practical problems, for example, in some optimization problems when large amount of high-fidelity data are required, or when the coefficients are very difficult to measure.

A more practical case is that only a limited number of high-fidelity observations are available, but sufficient low-fidelity approximations can be obtained via fast simulation. Multi-fidelity methods combine high-and low-fidelity data, and can be employed to efficiently achieve desired accuracy \cite{multif-surrogate, deepGaussian}. In \cite{comp-PINN}, the authors present a composite neural network (MPINNs) which can be trained utilizing multi-fidelity data. The architecture of the network consists of two neural networks, which aim to separately learn the nonlinear and linear correlations between the low-fidelity and the high-fidelity data, and then a physics-informed neural network to incorporate the physical laws with the multi-fidelity data. The proposed MPINNs can infer the quantities of interest given a few observations, and can identify the unknown parameters of the PDE by an inverse process. However, the data used for training are sampled from a fixed model/function, and once the model changed, the networks need to be retrained.

In this work, we aim to propose a general framework which can utilize sufficient low-fidelity data in conjunction with limited high-fidelity data from a class of models instead of a single model, and is able to provide an estimation of uncertainty in the prediction. In the mean time, our proposed framework should also take into account the available physical equations which govern the data. A neural process \cite{NP, ANP} is a probabilistic model that incorporates the idea from both neural network and Gaussian process. In a data-deficient case, it can learn distributions over functions using multiple data sets, which is achieved by some global latent variables in the model. This is superior compared with standard Gaussian process. These important features coincide with our purpose and are well suited to multi-fidelity problems where high and low-fidelity data are generated from multiple models. Under the idea of Bayesian modeling, it can encode prior beliefs about the correlation between multiple fidelities with a few observations and produce predictions with uncertainty. However, in the classical neural process, the physical information underlying the data is not involved. We will modify the loss function by imposing physical constraints to achieve a physical informed neural process.

We will take advantage of the natural property of neural process, adapt it to multi-fidelity data problems, and train the network with the help of physical constraints. Moreover, the constraints are added to the loss function using Lagrange multipliers. An algorithm to optimize the loss function is proposed to effectively training the parameters in the network on an ad hoc basis. By training the physical constrained neural process, one can obtain the correlation between the low- and high-fidelity data for a class of models instead of a fixed one. Applying the trained network, given a few low-fidelity observations from a new model, one can infer the high-fidelity data in the whole domain. Moreover, one can also identify the unknown parameters in the PDEs using our proposed method for inverse problems. 

The main contributions of our work include: (1) we propose a multi-fidelity physical constrained neural process (MFPC-Net) to learn correlations between multi-fidelity data which are generated from multiple models, (2) besides a few observations of multi-fidelity data, the network imposes physical laws of the underlying problem in loss functions, which can help the learning significantly in many applications, (3) with the help of the physical constraints, the prediction results from the network is much more accurate and reliable, (4) since the network learns from multiple data sets, the trained model can be employed as a robust surrogate, it is rapid for evaluation and generalization and produce predictions with uncertainty, (5) the proposed method can also be applied to inverse problems, and in this case the physical constraints are necessary in order to identify the unkown parameters in the PDE.

The paper is organized as follows, the main methods, architecture of the neural network and training algorithms are discussed in Section \ref{sec:main_methods}. The performance of the proposed methods is test on extensive numerical examples in Section \ref{sec:numerical}. We first test on one-dimensional pedagogical examples, and then on two-dimensional elliptic PDEs. For the elliptic PDE, we will consider both a forward and an inverse problem. The forward problem learns the high-fidelity solutions from random source terms. The inverse problems learn the solutions as well as identify the smooth or channelized permeability fields of the high-fidelity model. A conclusion is drawn in Section \ref{sec:conclusion}.

\section{Multi-fidelity neural process with physical constraints} \label{sec:main_methods}
We would like to model the relation between low and high-fidelity data under the idea of neural process with physical constraints. A comprehensive correlation between multi-fidelity models $y^H$ and $y^L$ can be represented as 
\begin{equation}
	y^H(x) = \mathcal{G}(y^L(x)) + \delta(x)
\end{equation}
where $\mathcal{G}$ in general is a nonlinear function which maps the low-fidelity data to high-fidelity ones, and $\delta(x)$ is space dependent and models the bias between low- and high-fidelity.

The architecture of the proposed multi-fidelity network will take into account data obtained from a class of low-and high-fidelity models, where we assume the observations in high-fidelity data are very limited. To be specific, consider a set of models (for example, a parametric partial differential equation which contains some random variables) instead of a single fixed model. Let $y^L(x)$ and $y^H(x)$ be low and high-fidelity data (for example, solutions of the parametric PDE with different fidelity of source terms of coefficients) obtained from these models. Suppose the low fidelity data $y^L(x)$ can be simulated in the entire computational domain, while the information of $y^H(x)$ can only be observed/measured in limited locations inside the domain. We would like to learn the correlations between this set of high and low fidelity models as a stochastic process with machine learning techniques. We remark that the data are sampled at a global level, where each sample is a vector contains several context/target points sampled from a function, rather than at a single local evaluation point on a function.That is, for each of the training sample, the input contains $m$ data pairs $(x_i, y^L(x_i)), i=1,\cdots, m$, and the output contains the corresponding values $y^H(x_i)$. Here,  $x_i$s are selected locations where one can acquire high fidelity data $y^H$ during the training. This will result in multiple data sets and thus is suitable for the application of neural process. The data is then divided into context set and target set, and model the conditional probability of the target given the context. The neural process will learn to adapt their priors to given data by minimizing the evidence lower bound (ELBO). The details of how neural process works will be discussed in the next section.

After training, the network will be capable of rapid generalization to new observations and can estimate the uncertainty in their predictions. That is, given context points from a new function, one can obtain the predictions on all other target points of this function.  Moreover, other than directly using the ELBO loss which is purely data driven, we will incorporate the physical information in the loss function as constraints, which can improve the learning of physical quantities of interest.

\subsection{Neural Process: in general}

Neural process (NP) \cite{NP, ANP} is a class of neural latent variable models which combines the idea of neural network and Gaussian process. NP models a stochastic process, which has the ability to represent a distribution over functions rather than a single function.

Consider a random process $F:X\rightarrow Y$. Let $(\textbf{x}_i, \textbf{y}_i)$, $i=1,\cdots, N$ be a set of observations. Split them into two sets, namely, the context points $(\textbf{x}_j, \textbf{y}_j), \; j = 1\cdots n$, and the target points $(\textbf{x}_t, \textbf{y}_t), \; t = n+1 \cdots N$. One aims to model the target set conditioned on the context. 

Following the concept from variational auto-encoders \cite{VAE}, let $z$ be the global latent variable such that $F(\textbf{x}) = g(\textbf{x},z)$. 
Here $h$ is deterministic and can be learned by the neural network. 
The global latent variables $z$ account for the uncertainty in the predictions.
One aims to maximize the marginal likelihoods of $(F(\textbf{x}_1),\cdots, F(\textbf{x}_N))$ under the generative process 
\begin{equation}
	p(\textbf{y}_{1:N}|\textbf{x}_{1:N}, z) = \int p(z) \prod_{i=1}^N \mathcal{N} (\textbf{y}_i | g(x_i, z), \sigma^2) dz
\end{equation}
The latent variable captures the global uncertainty when learning $F$.

The generation and inference work as follows:
\begin{itemize}
	\item[1.] Take the context points $(\textbf{x}_j, \textbf{y}_j)$ into an encoder $h$, which will produce a latent representation $r_j = h(\textbf{x}_j,\textbf{y}_j)$.
	\item[2.] Apply an aggregator $a$ on $r_j$, for example, summarize $r_j$ by $a$, that is, $r = a(r_j) =\frac{1}{n} \sum_{j=1}^n r_j$. 
	\item[3.] Create the distribution of latent variable $z$ parameterized by $r$, i.e. $p(z| x_{1:n},  y_{1:n}) = \mathcal{N} (\mu_z(r), \sigma_z^2(r))$. 
	\item[4.] Apply a decoder $g$ on a concatenation of sampled latent variable $z$ and the target point $\textbf{x}_t$, to obtain the predictions $\tilde{\textbf{y}}^t = g(\textbf{x}_t, z)$ which contains the mean and standard deviation information about the prediction.
	\item[5.] For the inference, apply the same encoder $h$ on both the context $(\textbf{x}_i, \textbf{y}_i), \; j=1:n$, and the target pairs $(\textbf{x}_i, {\textbf{y}}_i), \; j=n+1:N$, and then apply the aggregator $a$ to obtain $r$, and map $r$ to $\mu_z(r), \sigma_z^2(r)$.   Then the approximation posterior $q(z|\cdot ) =  \mathcal{N} (\mu_z(r), \sigma_z^2(r))$.
	\item[6.] Minimize the loss function involving ELBO. The variational lower bound is
	\begin{equation}\label{eq:ELBO}
		\log p(\tilde{\textbf{y}}_{1:N}|\textbf{x}_{1:N}) \geq \mathbb{E}_{q(z|\textbf{x}_{1:N}, \textbf{y}_{1:N} )}\large[ \sum_{i=n+1} ^N \log p({\textbf{y}}_i|z,\textbf{x}_i) + \log \frac{q(z|\textbf{x}_{1:n},\textbf{y}_{1:n} )}{q(z|\textbf{x}_{1:N},{\textbf{y}}_{1:N} )} \large]
	\end{equation} 
	where $q(z|\textbf{x}_{1:N},\textbf{y}_{1:N} )$ is a variational posterior of the latent variable $z$.
\end{itemize}

\subsection{Multi-fidelity models and physical constraint}
In our case, we are interested in the relation between the low and high fidelity data. For the learning, we adapt the input to be $\textbf{x} := (x, y^L(x))$ and the output to be $\textbf{y} = y^H(x)$. We remark that the input/output pairs will then be divided into context sets and target sets as described in the neural process in the training. 

For the loss function, we introduce additional physical constrained terms instead of simply minimizing the negative ELBO in \eqref{eq:ELBO}. In the standard negative ELBO approach, the loss consists of negative log likelihood of the target points, as well as the KL divergence between the variational posterior conditioned on context points and target points. It can be interpreted as minimizing the KL divergence subject to the reconstruction error \cite{GECO}. In this work, we would like to generalize the loss to minimizing the KL divergence and impose a set of constraints $\mathcal{C} (y^H_{1:N}, g(z))$ via Lagrange multipliers. The constraints typically involve the available physical law of the underlying problem.

\begin{equation}\label{eq:geco_loss}
	\mathcal{L}_{{\boldsymbol{\lambda}}; \theta} = \text{ELBO}  + {\boldsymbol{\lambda}}^T  \mathbb{E}_{q(z|x_{1:N}, y^L_{1:N},y^H_{1:N} )} [\mathcal{C} (y^H_{1:N}, g(x_{1:N}, y^L_{1:N}, z))]
\end{equation}
where 
$$\mathcal{C} (y^H_{1:N}, g(x_{1:N}, y^L_{1:N}, z)) = \begin{bmatrix}
||y^H - y^{H*}|| \\
||f^H - f^{H*}||
\end{bmatrix} - \boldsymbol{\tau} $$
$g$ is the decoder, $\boldsymbol{\lambda}$ is the Lagrange multiplier, and $\theta$ denotes all the parameters in the neural network, $\boldsymbol{\tau}$ is a given threshold. $y^{H*}$ is the mean approximation of $y^{H}$ obtained from the neural network decoder. $f^H  = \mathcal{L}(x, y^H)$, which can be directly measured/observed, and $f^{H*}= \mathcal{L} (x, y^{H*})$ where $\mathcal{L}$ denotes the physical equation/law. We remark that, here we are only given the data at limited locations, and the physical equation are also constrained at selected locations instead of all locations in the entire computational domain. Then the physical constraints $||y^H -  y^{H*}|| = \frac{1}{N_s} \sum_{i=1}^{N} ||y_i^H - y_i^{H*} ||_2$ and $||f^H - f^{H*}|| = \frac{1}{N_f} \sum_{i=1}^{N_s} ||f_i^H - f_i^{H*} ||_2$, where $N_s$ is the number of samples. The term $f^H  = \mathcal{L}(x, y^H)$ can be the original PDE, or a weak form of the equation.

\begin{figure}
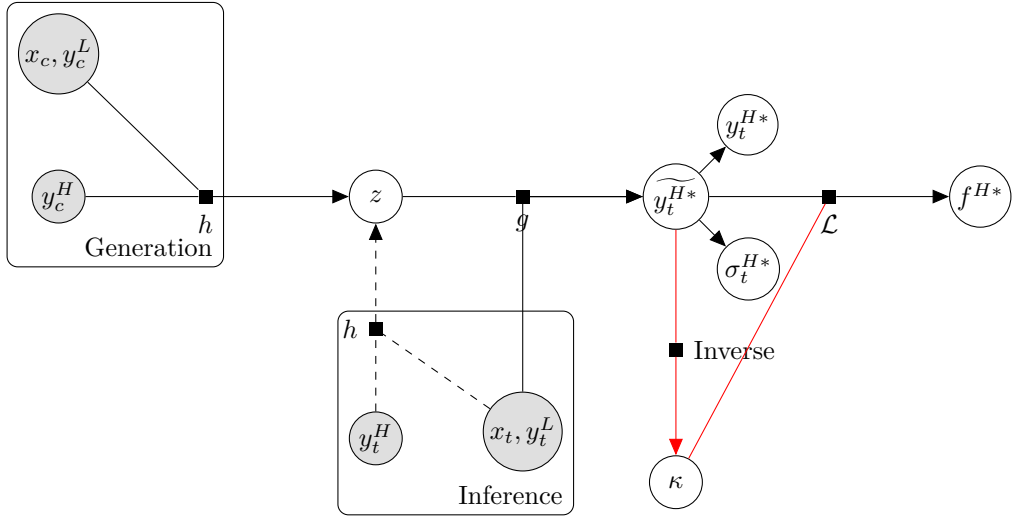

	\begin{center}
		\tikz{ %
			\node[obs] (yhc) {$y^H_c$}; 
			\node[obs, above=of yhc] (xcylc) {$x_c, y^L_c$}; 
			\node[latent, right =3.5cm of yhc] (z) {$z$} ; %
			\factor[right =1.5cm of yhc] {h} {below:  $h$ } {yhc, xcylc} {z} ; 
			\factor[right =1.5cm of z] {g} {below:  $g$ } {} {} ; 
			\node[latent, right =1.5cm of g] (yt) {$\widetilde{y_t^{H*}}$} ; %
			\node[latent, above right =0.5cm of yt] (my) {$y_t^{H*}$} ; 
			\node[latent, below right =0.5 cm of yt] (sy) {$\sigma_t^{H*}$} ; 
			\factoredge {z} {g} {yt} ; 
			\factoredge  {g}{yt} {my, sy}; 
			
			\node[obs, below =2.5 cm of g] (xtylt) {$x_t, y^L_t$}; 
			\node[obs, below=2.5 cm of z] (yht) {$ y^H_t$}; 
			\factor[above =1 cm of yht] {h2} {left:  $h$ } {} {} ; 
			\factoredge[dashed]{yht, xtylt}{h2} {z}
			\factoredge{xtylt}{g} {yt}
			
			\factor[right =1.5 cm of yt] {L} {below:  $\mathcal{L}$ } {} {} 
			\node[latent, right =1.5cm of L] (ft) {$f^{H*}$} ; 
			\factoredge{yt}{L} {ft}
			
			\node[latent, below =3 cm of yt] (kappa) {$\kappa$} ; 
			\factor[below =1.5 cm of yt] {inv} {right: Inverse }{}{}
			\factoredge[red]{yt} {inv} {kappa} 
			\factoredge[red]{kappa}{L} {}
			
			\plate {plate1} { %
				(xcylc)(yhc)(h) %
			} {Generation} ; %
			
			\plate {plate2} { %
				(xtylt)(yht)(h2) %
			} {Inference} ; %
		}
	\end{center} \caption{Illustration of the proposed network. $h$: encoder. $g$: decoder. $\mathcal{L}$: the physical equation. $\mu_y$: mean of the prediction at target points, $\sigma_y$: standard deviation of the prediction at target points. $k$: quantities of interest in the inverse problem. The black-line part represents the net for forward problems. With additional red-line part, it represents the net for inverse problems. } \label{fig:network}
\end{figure}

The architecture of the network is shown in Figure \ref{fig:network}. In the training stage, we are given the observations of $\Big((x_c, y^L_c), y^H_c\Big)$, as well as $\Big((x_t, y^L_t), y^H_t\Big)$, where the subscript $c$ denotes context, $t$ denotes target.The context sets are first feed in the latent encoder $h$ and aggregator to obtain the variational prior distribution of global latent variable $z$. Then the target sets come to the network as well for inference purposes, to provide a variational posterior distribution of $z$. The KL divergence of these two and the log probability of $y_t^{H}$ contribute to parts of the loss function. Next, the target input $(x_t, y^L_t)$ only is passed to the decoder to get the predictions $\tilde{y_t^{H*}}$ of $y_t^H$, which contains both the mean $y_t^{H*}$ and standard deviation $\sigma_t^{H*}$ of the approximation. The physical law $\mathcal{L}$ is then applied on the mean of the approximation $y_t^{H*}$, to generate the right hand side of the equation $f^{H*}$. The relative mean of the reconstruction error $||y^H_t-y_t^{H*}||$ and  feature reconstruction error $||f^H-f^{H*}||$ will contribute to the loss function as Lagrange multipliers. In the testing stage, one will not have the observations for $y_t^H$. With only $x_t, y^L_t$ as the input to the decoder, one can then get the approximations for $y_t^H$ at any interested target locations. This is for the forward problem. The rest of the diagram is for inverse problems and will be discussed in the next section.

The new loss function \eqref{eq:geco_loss} is then optimized with respect to the neural process parameters $\theta$ and $\boldsymbol{\lambda}$. Following the idea from \cite{GECO}, $\theta$ is minimized using the stochastic gradient descent, and the Lagrange multipliers are maximized following a moving average of the constraints, as described in Algorithm \ref{alg:GECO}. 

Typically, at the early training stage, the optimization problem will shrink the search area for the model parameters by making them satisfy the constraints, and then it minimizes ELBO while preserving the validity of the current parameter solutions. 

The idea is as shown in Algorithm \ref{alg:GECO}.
\begin{algorithm}
	\caption{Physical Constrained Neural Process with Multi-fidelity data (MFPC-Net)}\label{alg:GECO}
	\begin{algorithmic}[1]
		\Procedure{MFPC-Net}{  {it}$=0$, $\theta_0, \lambda_0 = 1, {n_{it}}, \alpha = 0.9$}
		\While{${it} < {n_{it}} $}
		\State Read data batch $\Big( (x_{1:N}, y^L_{1:N}),  y^H_{1:N} \Big)$
		\State Divide into context sets $\Big( (x_{1:n}, y^L_{1:n}),  y^H_{1:n} \Big)$ and target sets $\Big( (x_{1:N}, y^L_{1:N}),  y^H_{1:N} \Big)$
		\State $\theta \gets \theta_0$, $\boldsymbol { \lambda} \gets \lambda_0$
		\State Apply latent encoder on context sets  to get variational prior
		\State Apply latent encoder on target sets  to get variational posterior
		\State Sample $z$ from variational posterior $q(z|(x_{1:n}, y^L_{1:n}),y^H_{1:n} )$
		\State Apply decoder $g(z)$ on $(x_{1:N}, y^L_{1:N})$ to obtain approximation  $y^{H*}$
		\State Compute the average of the constraints $C^{{it}} \gets \mathcal{C} (y^H_{1:N}, y^{H*})$ 
		\If {${it} == 0$}
		\State $C_{ma}^{{it}} \gets  C^{{it}}$
		\Else
		\State $C_{ma}^{{it}} \gets  \alpha C_{ma}^{{it}-1} + (1-\alpha) C^{{it}}$
		\EndIf
		\State Compute loss from \eqref{eq:geco_loss} 
		\State $\theta_0 \gets \argmin (L_{\boldsymbol { \lambda } })$ using appropriate stochastic gradient descent algorithm
		\State $\lambda_0 \gets \boldsymbol{\lambda} \exp(C_{ma}^{{it}} )$ 
		\State  ${it}  \gets  {it}+1$
		\EndWhile
		\EndProcedure
	\end{algorithmic}
\end{algorithm}

\subsection{Inverse problem}

We can further extend the proposed method to some inverse applications. In some engineering problems, the coefficients in the operator $\mathcal{L}$ are not known. For example, it might be harder to measure the permeability coefficients $\kappa$ than to measure the solution pressure $u$ in the differential equation
\begin{equation*}
-\text{div} (\kappa \nabla u ) = f.
\end{equation*}
 It would be interesting to learn the coefficients from the solution of the equation. However, the measurements of high fidelity solution $u^H$ are also expensive, one may only obtain observations of $u^H$ is a limited number of locations. It is common to have enormous low-fidelity simulations in conjunction with the limited high-fidelity observed data together. One then aims to infer the high fidelity solution or the underlying true permeability. We can achieve this idea by the proposed network Figure \ref{fig:network}, where the inverse steps are denoted by the red lines. That is, we can further add a few layers in the network to model from predicted high fidelity solution $\tilde{y^{H*}}$ to the coefficients $\kappa$. Moreover, to impose physical constraints, the coefficient $\kappa$ is needed. Now, we can pass the predicted $\kappa$ into equation $\mathcal{L}$ to further realize the constraints. We remark in this case, the loss function also includes the error between the true and predicted $\kappa$. That is,

\begin{equation}\label{eq:geco_loss_inverse}
	\mathcal{L}_{{\boldsymbol{\lambda}}; \theta} = \text{ELBO}  + {\boldsymbol{\lambda}}^T \begin{bmatrix}
		||y^H - y^{H*}|| \\
		||F^H - F^{H*}||\\
		||\kappa - \kappa^{*}||
	\end{bmatrix}- \boldsymbol{\tau}
\end{equation}
where $\kappa^*$ is the prediction of the coefficient $\kappa$ from the network for the inverse problem. In addition, for the constraint on the data $F^H$, the predicted values $F^{H*} = \mathcal{L}(x, y^{H*}; \kappa*)$ need to be computed using $\kappa^*$ since the true $\kappa$ is unknown quantity of interest. We remark that, in our example on the two dimensional elliptic problem, we use the weak formulation to impose this constraint, that is  
\[
F_i^H -  F_i^{H*} = \int_{\Omega} f_i^H v - \int_{\Omega} \kappa^* \nabla y_i^H \nabla v = F_i^H - A(\kappa^*)  y_i^H
\]
where $A(\kappa^)$ is the stiffness matrix. Then $||F_i^H - F_i^{H*}||_2/||F_i^H||_2$ denotes the relative residual of the linear system.

\section{Numerical examples}\label{sec:numerical}
In this section, we are going to provide some typical examples to justify our proposed methods.
\subsection{Continuous function with nonlinear correlation}
To test the performance of the presented method, we first consider the following problem.
\begin{align}
	y^L(x) &= \sin {\alpha \pi x},\\
	y^H(x) &= (x- \beta) (y^L)^2  
\end{align}
where, $x\in [0, 1]$, $\alpha \sim U([2,5])$, $\beta \sim U([-4, 4])$ are random variables. Here the two levels of fidelity have nonlinear correlations. Moreover, since $\beta$ is also a random variable, for given $\alpha$, the relationship between every pair of high and low fidelity functions is not deterministic. We aim to learn this stochastic relation using neural process. 

For testing purposes, in this example, we assume the second derivative of $y^H(x)$ can be measured/observed at some locations for the training, i.e. 
\begin{align}
	f^H(x) &= (y^H){''}(x) = 2 ( \alpha \pi)^2  (x- \beta) \cos(2 \alpha \pi x ) + 2 \alpha \pi \sin(2 \alpha \pi x)
\end{align}
Then we can strongly impose constraints $||f^H - f^{H*}||$ in the loss function as described in the previous section.

\subsubsection{With a fixed number of total samples, and fixed number of data points}
First, we choose an initial of $200$ samples and then actively add a batch of samples according to their predictive variances. For the active learning algorithm, please refer to Appendix \ref{sec:append_a}. The batch size is chosen to be $50$. In the last epoch, the number of training samples will be $4000$. In this example, there are 4 layers and 64 neurons per layer adopted for both encoder and decoder in the neural network. 

\begin{figure}[!htb]
	\begin{center}
		\includegraphics[scale=0.4]{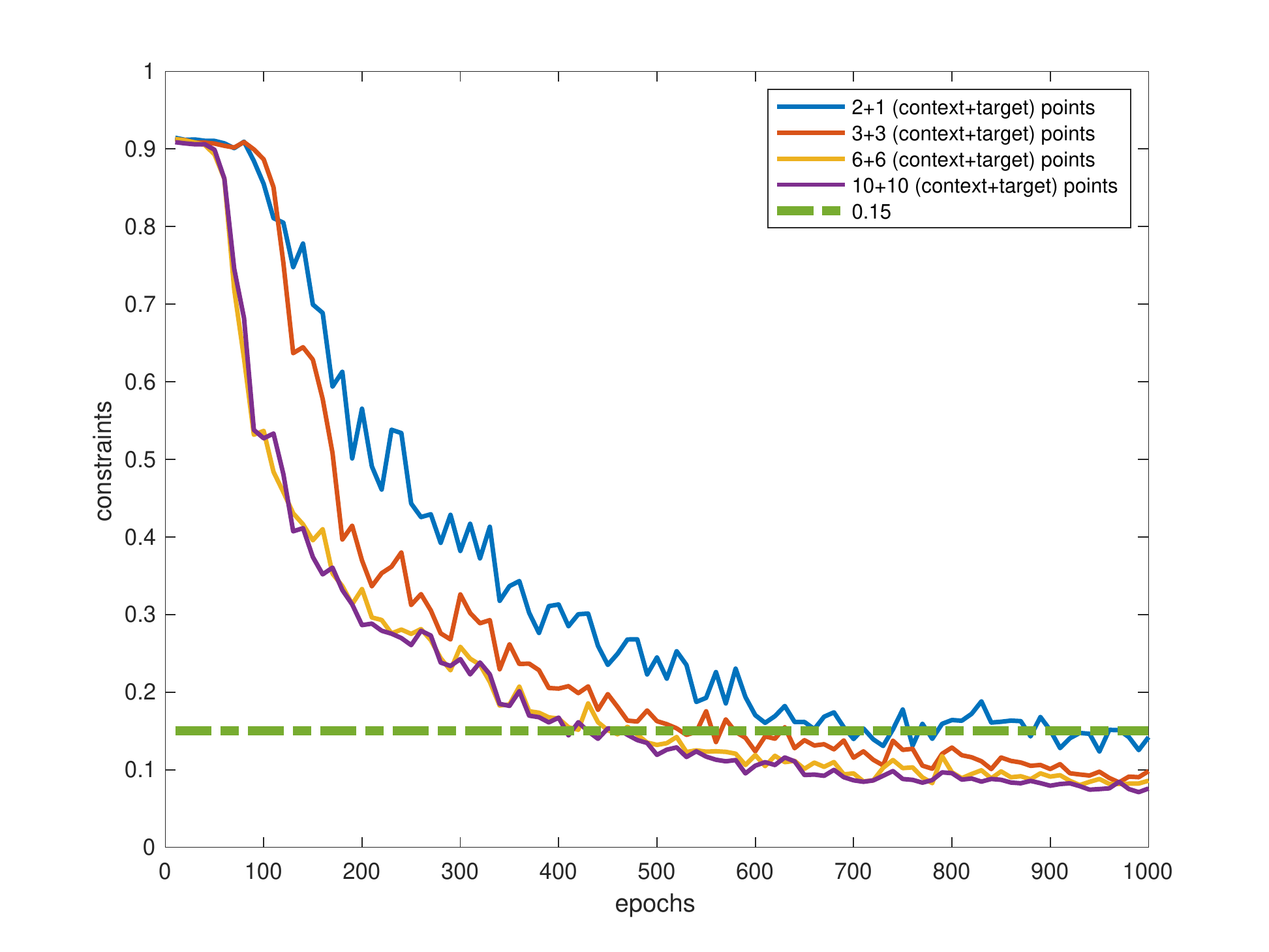}
	\end{center}
	\caption{The physical constraints value during the training, when taking (1) 2 context points + 1 target points, (2) 3 context points + 3 target points, (3) 6 context points + 6 target points, and (4) 10 context points + 10 target points.}  \label{fig:ex1-constraints}
\end{figure}

During the training, the context/target points are chosen to be evenly distributed points in the interval $x\in [0, 1]$. We use different number of context and target points to train the network, and the experiments show that only a few points will provide promising results. As for the number of data points ($N_f$) with respect to $f^H$, we first fixed it to be $N_f = 20$, and investigate the performance of the proposed method when taking different number of context points in $y^H$.
In Figure \ref{fig:ex1-constraints}, we present the average value of the constraint $||f^H - f^{H*}||$ among all training samples during the training. In this example, we set the threshold $ \boldsymbol{\tau}_2 = 0.15$. It shows that within 1000 epochs, the constraints threshold can be achieved with a very small number of context and target points. As we increase the context and target points, the values convergence faster. The CPU time in each training epoch is around 0.027 second.

\begin{figure}[!htp]
	\begin{center}
		\begin{subfigure}[t]{1.0\textwidth}
			\centering
			\includegraphics[scale=0.23]{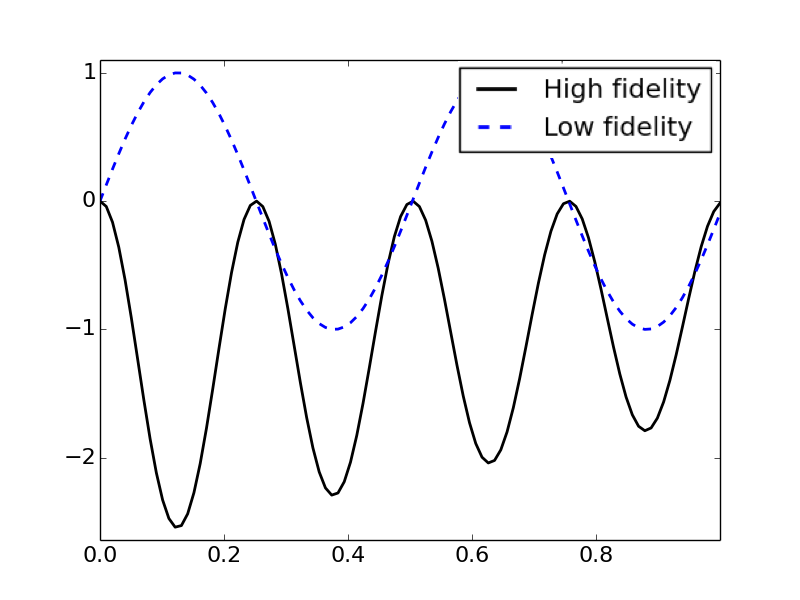}
			\includegraphics[scale=0.23]{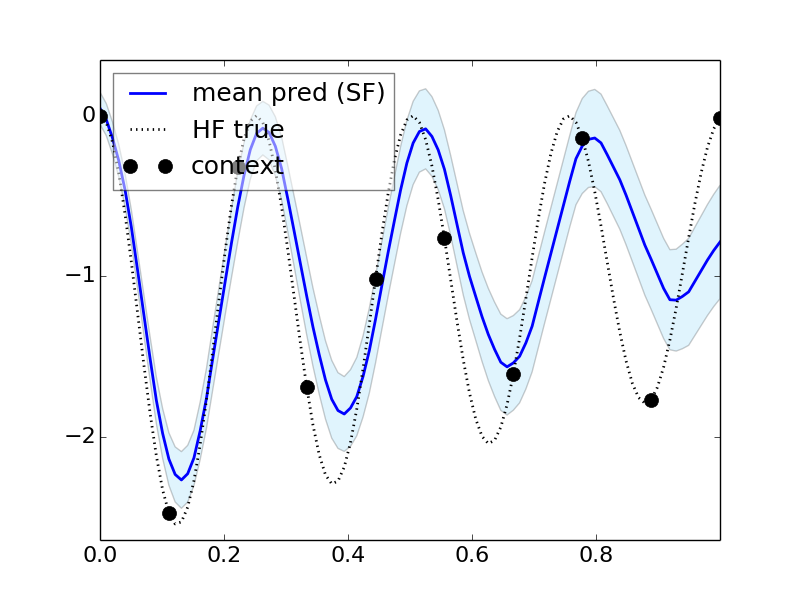}\\
			\includegraphics[scale=0.23]{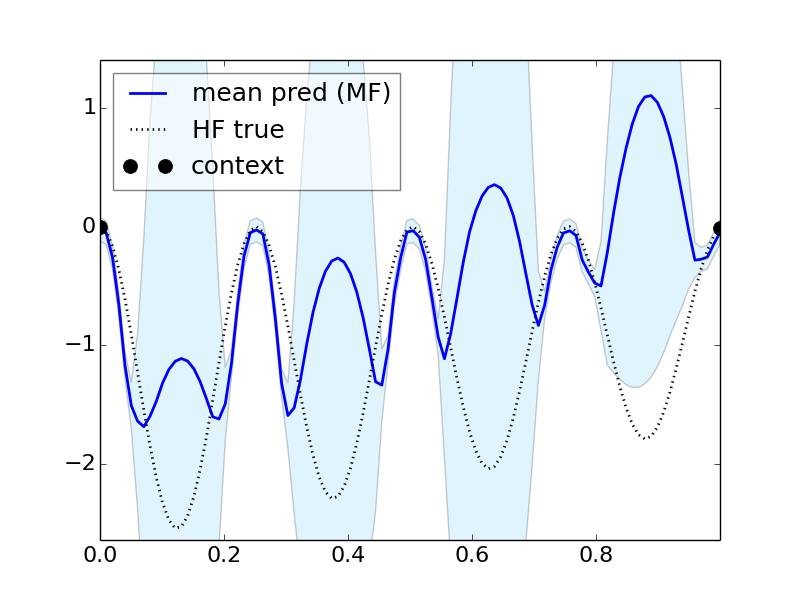}
			\includegraphics[scale=0.23]{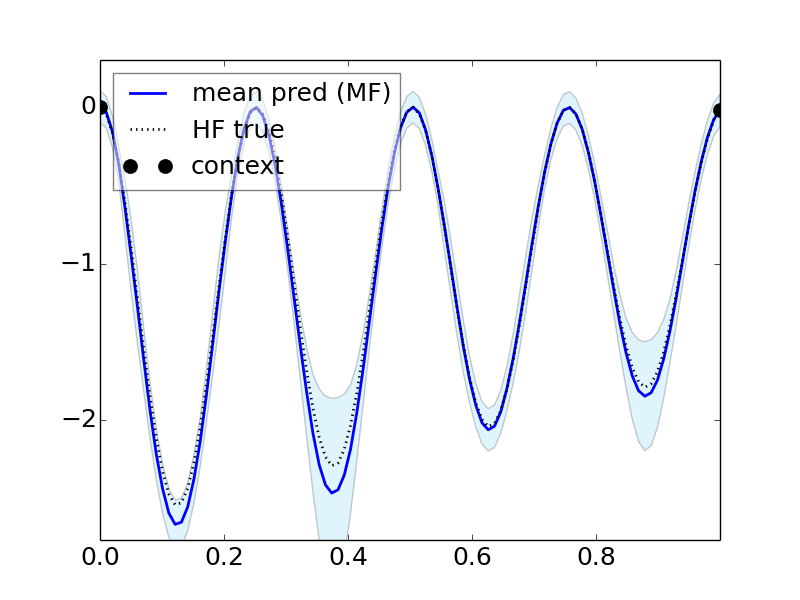}\\
			\caption{test sample 1}
		\end{subfigure}
		\begin{subfigure}[t]{1.0\textwidth}
			\centering
			\includegraphics[scale=0.23]{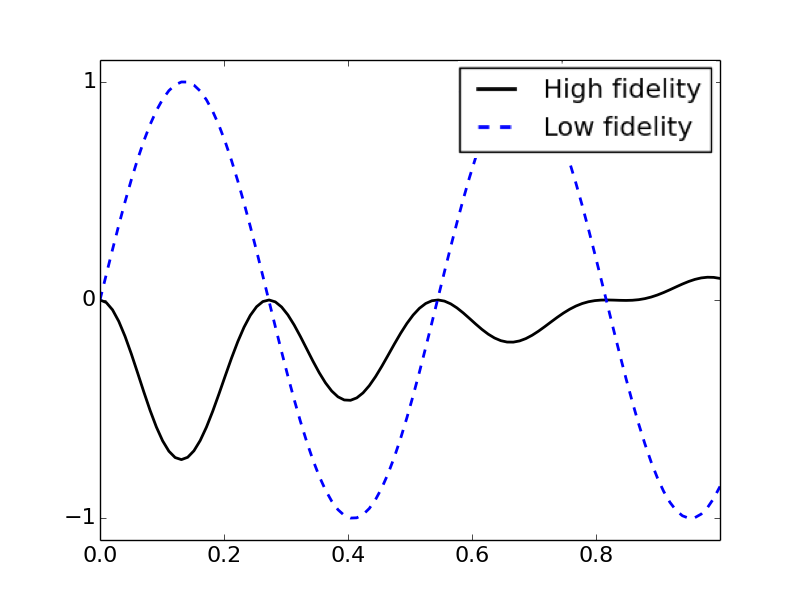}
			\includegraphics[scale=0.23]{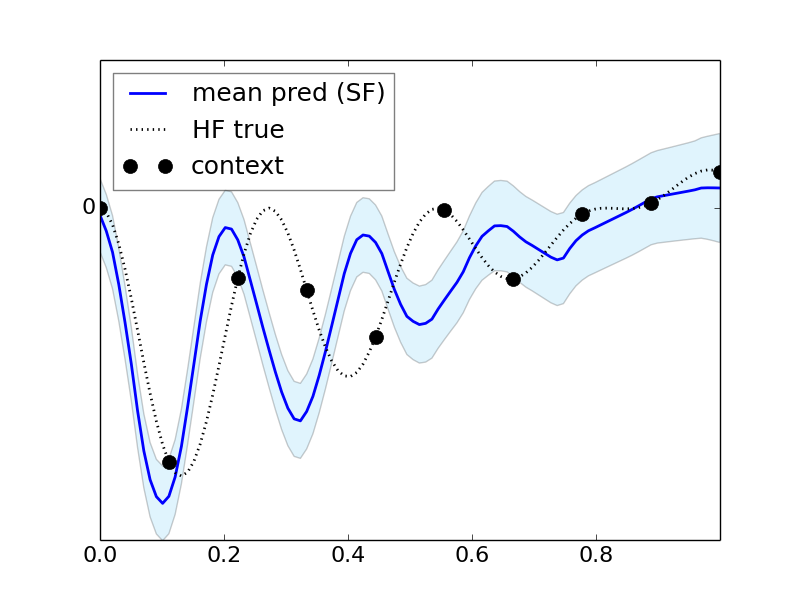}\\
			\includegraphics[scale=0.23]{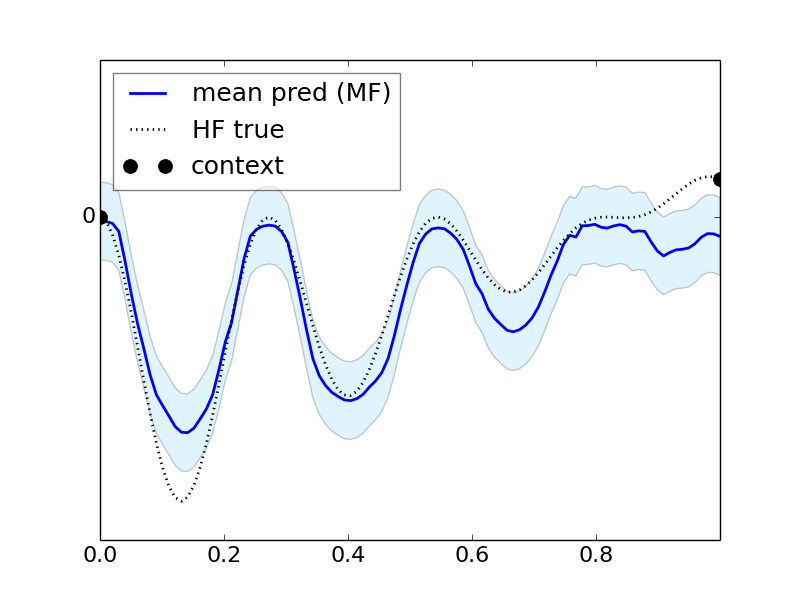}
			\includegraphics[scale=0.23]{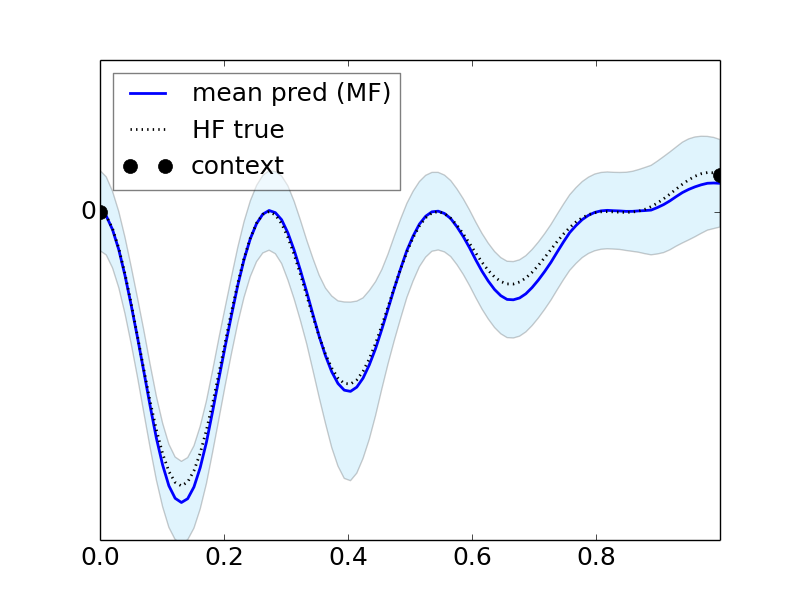}\\
			\caption{test sample 2}
		\end{subfigure}
		
		\begin{subfigure}[t]{1.0\textwidth}
			\centering
			\includegraphics[scale=0.23]{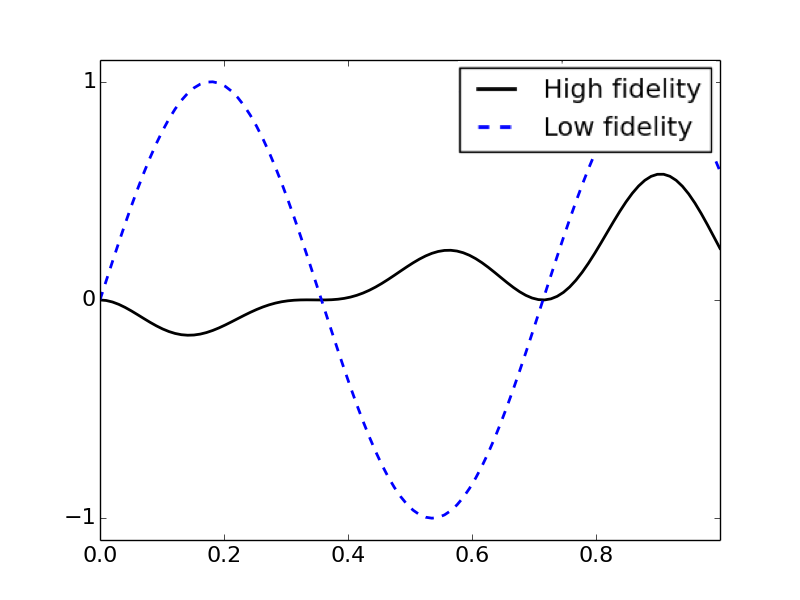}
			\includegraphics[scale=0.23]{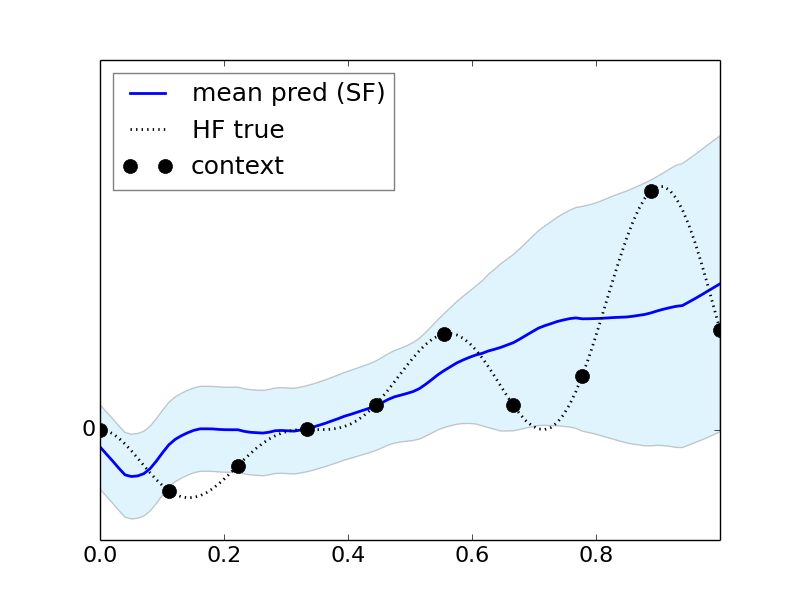}\\
			\includegraphics[scale=0.23]{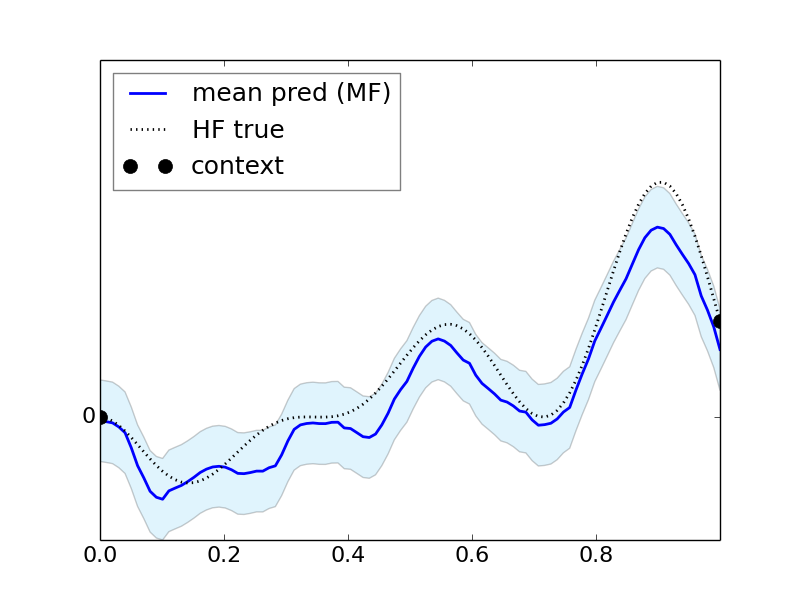}
			\includegraphics[scale=0.23]{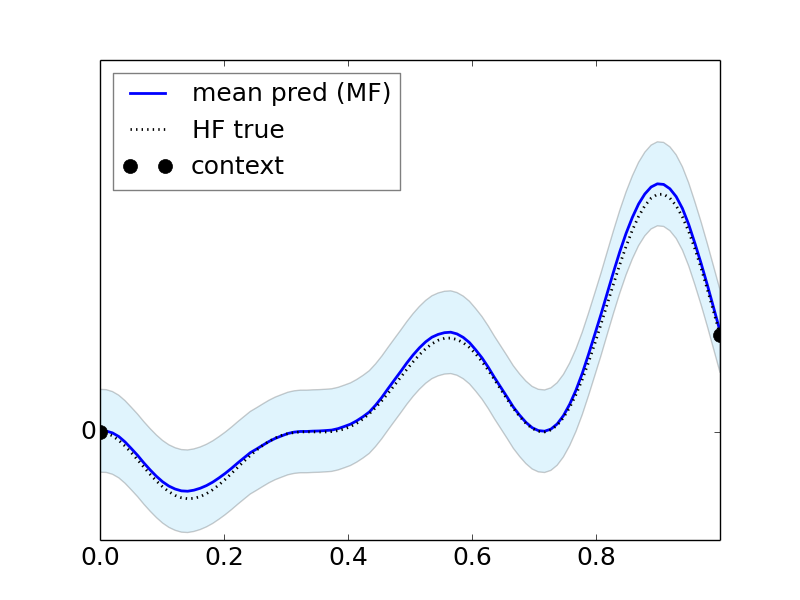}\\
			\caption{test sample 3}
		\end{subfigure}
	\end{center}
	\caption{Comparison of results using single fidelity data, using multi-fidelity data without physical constraints,  and using multi-fidelity data with physical constraints.  In each sub-figure, from left to right, top to bottom: (1) low and high fidelity data; (2) prediction using single fidelity data, with 10 context and target points; (3) prediction using multi-fidelity data, with 2 context and 1 target points, no physical constraints; (4) prediction using multi-fidelity data, with 2 context and 1 target points, with physical constraints. } \label{fig:ex1-samples}
\end{figure}

As for the validation/testing, when there are only two context points, i.e., $x=0, x=1$, we present the predictions for three test samples in Figure \ref{fig:ex1-samples} (each row corresponds to a different sample). The black markers in each subplot denotes in the context sets in the test. The dashed black curves denote the true high-fidelity data. The blue curves are the mean prediction for the high-fidelity function, and the blue shaded is the continuous error bar. The three subplots on the second column of Figure \ref{fig:ex1-samples} are obtained from the network with only single fidelity data, the third column is predictions from the network trained using multi-fidelity data but without physical constraints, and the three subplots on the fourth column of Figure \ref{fig:ex1-samples} are the results from the network using multi-fidelity data as well as constraints on the second derivative of the data $||f^H - f^{H*}||$. Here we assume there are $20$ points on $f^H $ can be observed in all cases.  We can see that with only single fidelity data, we need a lot more context points, but still get unreliable predictions. With multi-fidelity data, the predictions are good with only very few context points. What is more, there is a great improvement on the prediction when using physical constraints and training the network using Algorithm \ref{alg:GECO}.

\begin{figure}[!htp]
	\begin{center}
		\includegraphics[scale=0.4]{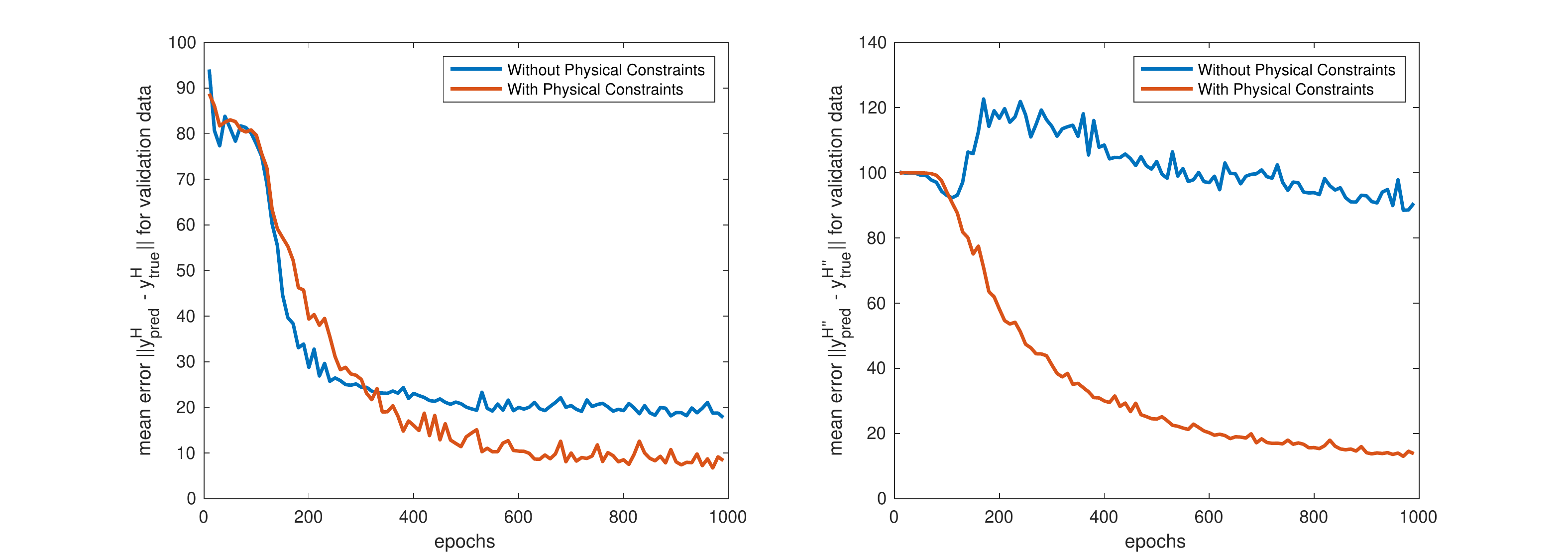}
		\includegraphics[scale=0.4]{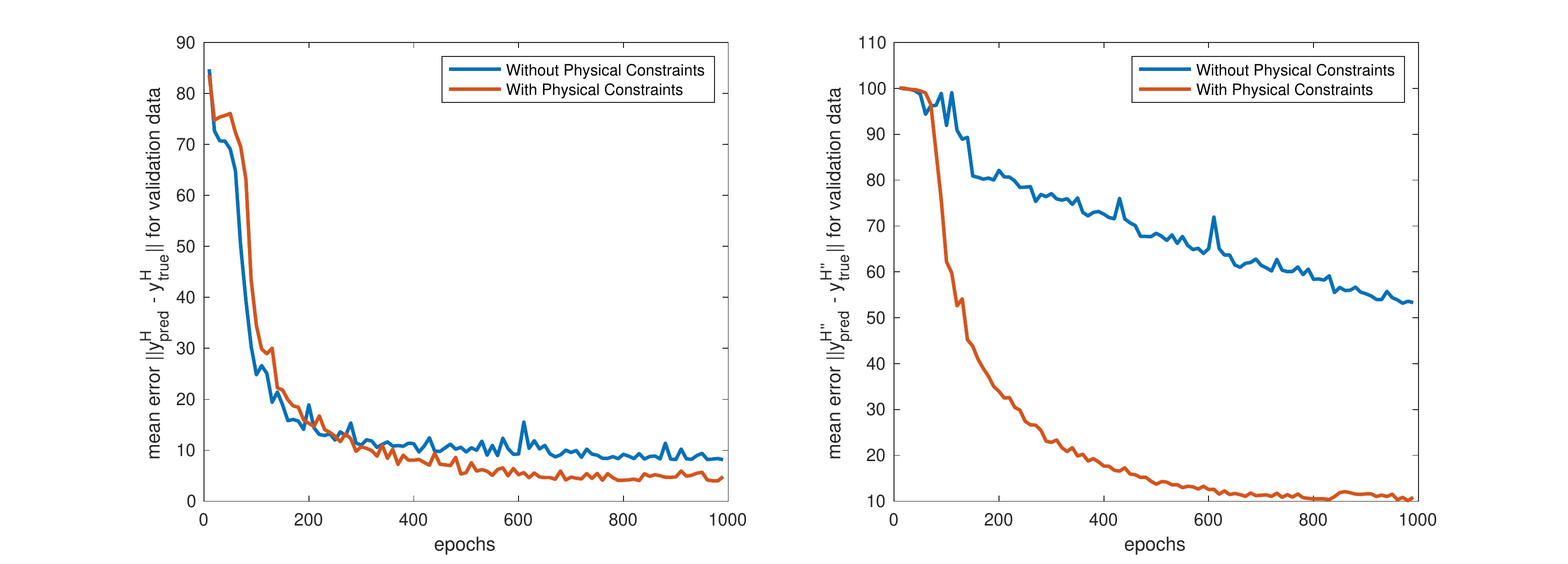}
		\includegraphics[scale=0.4]{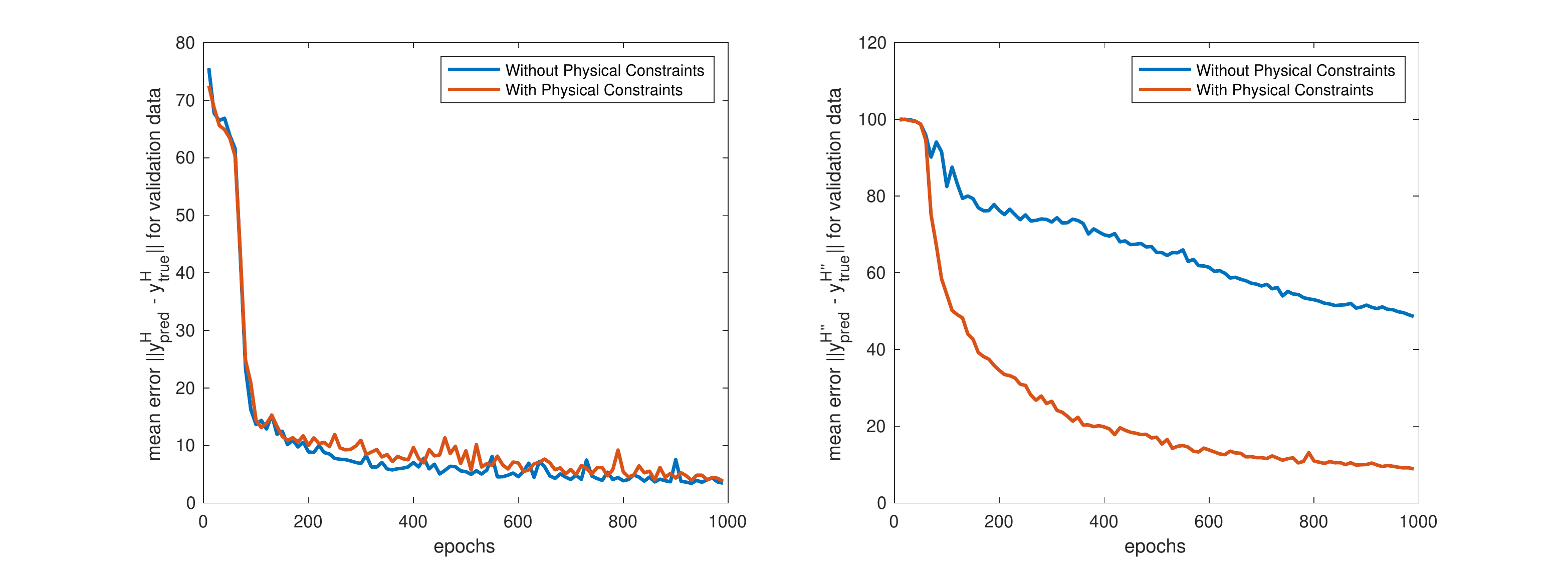}
		\caption{The mean errors for validation data, when taking (1) 2 context points (top subplot), (2) 3 context points (middle subplot), and (3) 6 context points (bottom subplot).}  \label{fig:ex1-contexts}
	\end{center}
\end{figure}

We then compare the mean errors of the testing samples when taking different number of context/target points, see Figure \ref{fig:ex1-contexts} for an illustration. We note that all errors are in percentage. We remark, in this figure, we still take a fixed number of data points $N_f = 20$, taking other fixed numbers $N_f$ will result in similar behavior. The blue curves and red curves correspond to the results without and with physical constraints, respectively. As we increase the number context points, both relative mean errors $||y^H_{\text{true}} -y^H_{\text{pred}}||$ and $||(y^H_{\text{true}})'' -(y^H_{\text{pred}})''||$ for validation/test samples decrease faster against epochs. With very small number of context points (subplot in the first row of Figure \ref{fig:ex1-contexts}), the pure data-driven approach has much larger errors, especially for the $||(y^H_{\text{true}})'' -(y^H_{\text{pred}})''||$ error, adding physical constraint can significantly reduce those errors. As we increase the context points (subplot in the second and third rows of Figure \ref{fig:ex1-contexts}), the pure data-driven approach can already have very good performance for $||y^H_{\text{true}} -y^H_{\text{pred}}||$ errors thanks to the power of neural process, while adding physical constraints can also help, especially for $||(y^H_{\text{true}})'' -(y^H_{\text{pred}})''||$ errors again.

\subsubsection{With a different number of total samples, and different number of data points}
\begin{figure}[!htp]
	\begin{center}
		\includegraphics[scale=0.35,height=6cm]{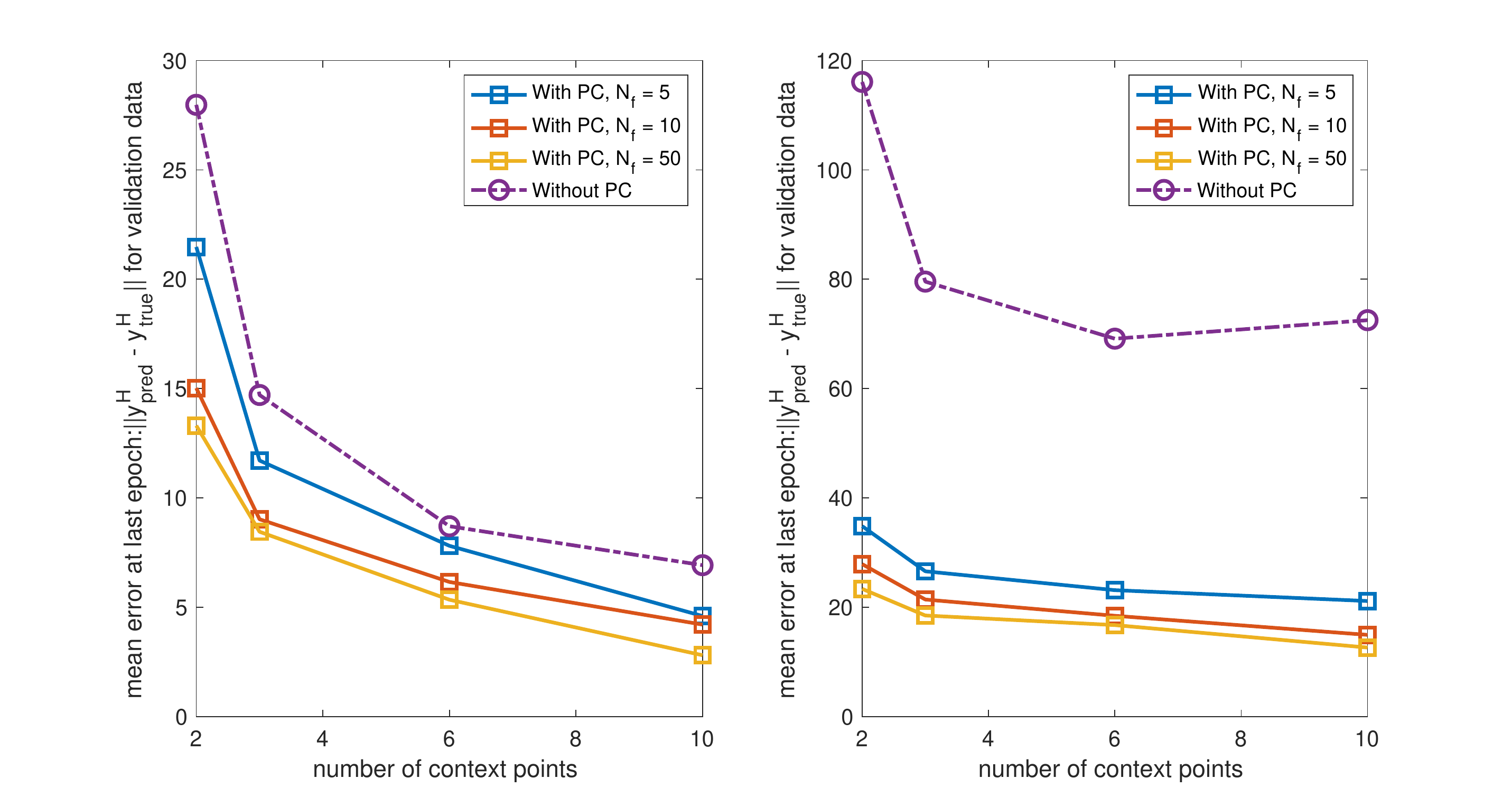}
		\includegraphics[scale=0.35,height=6cm]{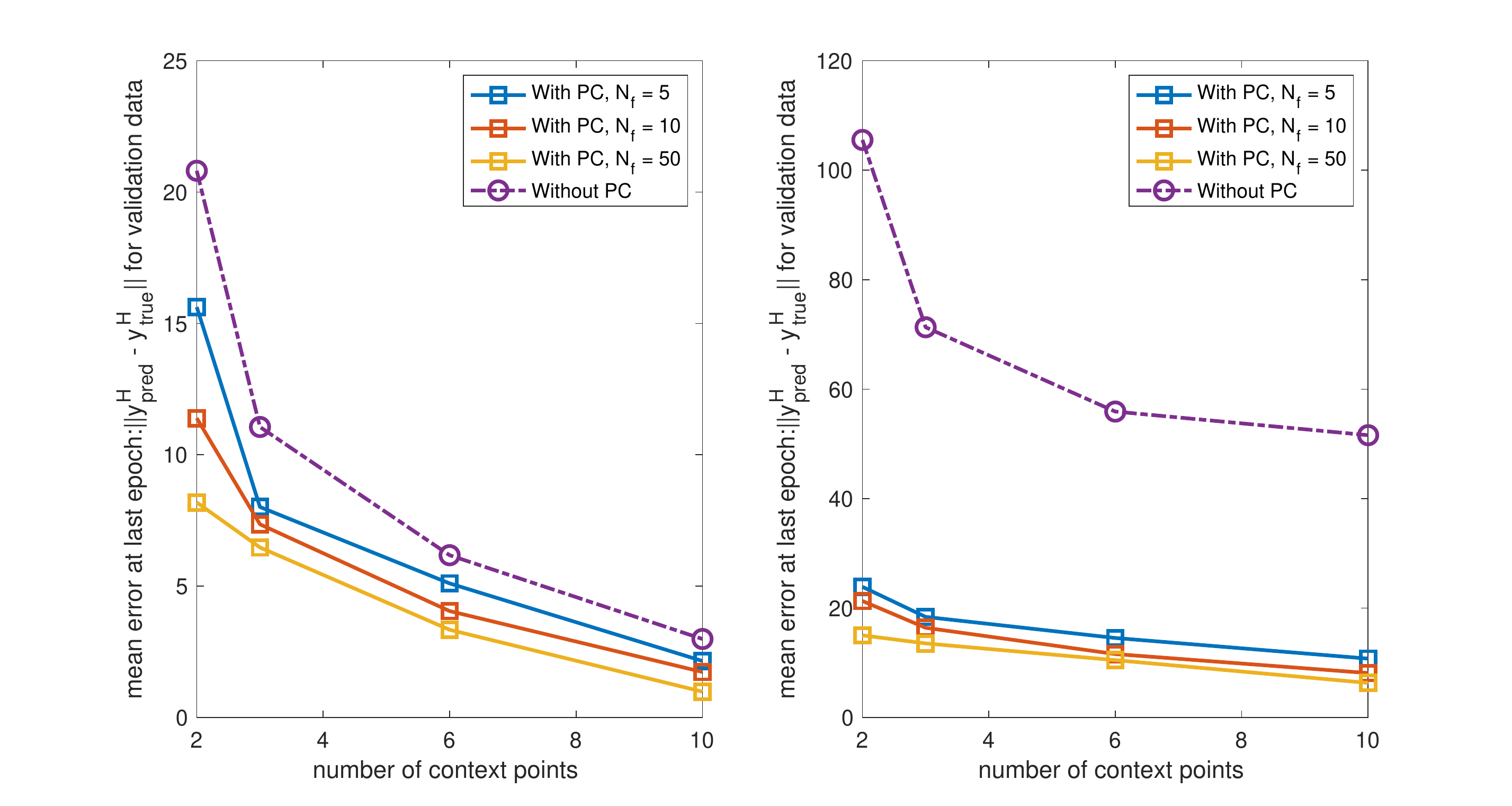}
		\includegraphics[scale=0.35,height=6cm]{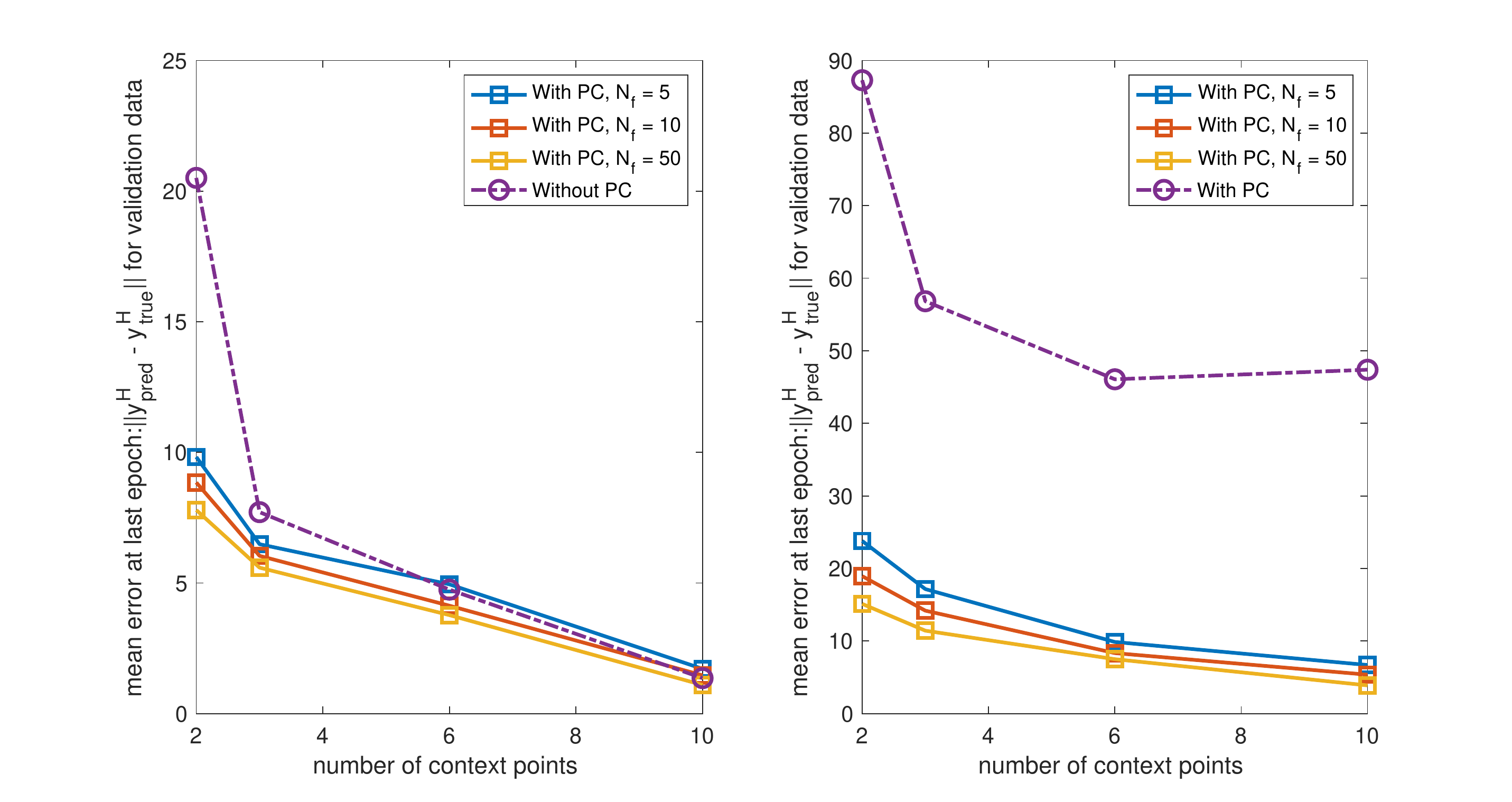}
		\caption{The mean errors for 200 validation samples at the last epoch when taking different number of context points and different data points. The context points refer to the number $n$ of the high fidelity data, and the data points refer to the number $N_f$ of the points $f^H$ in the physical constraints. (1) 200 training samples (top subplot), (2) 1000 training samples (middle subplot), and (3) 4000 training samples (bottom subplot). }\label{fig:ex1-lastepoch}
	\end{center}
\end{figure}

In the end, we also compare the performance of the proposed method when taking different numbers of training samples as well as different values of $N_f$ in Figure \ref{fig:ex1-lastepoch}.
With fixed number of training samples, as we increase $N_f$, the mean errors can also decrease in all cases. As we increase the number of training samples, the overall errors will decrease fast. When there are enough training samples, the contrast among the results with physical constraints and without physical constraints become less obvious for $||(y^H_{\text{true}}) -(y^H_{\text{pred}})||$, but we always observe a significant reduction in the relative mean errors $||(y^H_{\text{true}})'' -(y^H_{\text{pred}})''||$ when use physical constraints, as expected.This shows that, our method has good performances when we have limited number of training samples. 

\subsection{Elliptic PDE: forward problem}
In this section, we apply our proposed method to 2D elliptic differential equations. Let $\mu$ be some random parameters. Define the high fidelity model, which is to find $u(\mu) \in V_h$ such that 
\begin{equation} \label{eq:elliptic-f-high}
	a(u_H(\mu), v) = (f(\mu),v), \quad \quad \forall v \in V_h
\end{equation}
or
\begin{equation} \label{eq:elliptic-k-high}
	a(u_H(\mu), v; \mu) = (f,v), \quad \quad \forall v \in V_h
\end{equation}
where $a(u, v) = \int_{\Omega} \kappa  \nabla u \nabla v$, $u_H(\mu)$ is the fine scale solution to the PDE for an instantiation of the random parameter $\mu$, $V_h$ is the finite element space with piecewise linear basis functions. Here the random $\mu$ may lie in the source term as in \eqref{eq:elliptic-f-high} or lie in the diffusion coefficient $\kappa(\mu)$ as in \eqref{eq:elliptic-k-high}. We assume there are some observations of the true quantities of interest 
$$
\textbf{s}_i (\mu) = c (u_H(\mu)).
$$

As for the low fidelity models, we find $u_l(\mu)\in V_h$ such that
\begin{equation} \label{eq:elliptic-f-low}
	a(u_l(\mu), v) = (f_l(\mu),v), \quad \quad \forall v \in V_h
\end{equation}
or
\begin{equation}\label{eq:elliptic-k-low}
	a(u_L(\mu^0), v; \mu^0) = (f,v), \quad \quad \forall v \in V_h
\end{equation}
and the simulated quantity of interest satisfy 
$$
\tilde{\textbf{s}_i} (\mu) = c (u_L(\mu)).
$$
Here, we assume the source term in \eqref{eq:elliptic-f-low} lose some information compared with the true source in \eqref{eq:elliptic-f-high}, but the random variable $\mu$ are the same in each term. In the second case \eqref{eq:elliptic-k-low}, we assume the permeability field $\kappa(\mu^0)$ has lower fidelity compared with the true permeability in \eqref{eq:elliptic-k-high}. We are interested in a forward problem for \eqref{eq:elliptic-f-low}/\eqref{eq:elliptic-f-high}, and an inverse problem for 
\eqref{eq:elliptic-k-low}/\eqref{eq:elliptic-k-high}.



\subsubsection{High and low fidelity models with different source terms}

We first consider the elliptic equation with low and high fidelity source terms, 
\begin{equation} \label{eq:elliptic-f}
	-div (\kappa \nabla u ) = f(\mu)  \;\; \text{in} \;\; \Omega
\end{equation}
with Dirichlet boundary condition $u = 0$ on $\partial \Omega$, where $\mu$ are some random parameters, $\kappa= 1$ across the domain in this example.

For the high fidelity models, we choose the source term 
\begin{equation}
	f_H(x;\mu) = \sin(\mu_1 \pi x) \sin(\mu_2 \pi y) +\sin(2 \mu_1 \pi x) \sin(2\mu_2 \pi y) 
\end{equation}
where $ \mu = (\mu_1, \mu_2)^T$,  $\mu_i \sim N(0,1)$ for $i=1,2$.

For the low fidelity model, we take 
\begin{equation}
	f_L(x;\mu) = \sin(\mu_1 \pi x) \sin(\mu_2 \pi y) 
\end{equation}

The observational functional is chosen to be $c_i(v) = \int_{\Omega} \delta(x_i) v dx$, $i = 1, \cdots, N$. 

We apply the proposed algorithm to estimate the high fidelity solution of the equation \eqref{eq:elliptic-f}. There are 1000 samples for training, and 200 samples for testing. Suppose $f_L$ is known in the entire domain, thus the low fidelity solution $u_L$ can be solved using finite element simulation. On the other hand, $f_H$ as well as $u_H$ can only be measured in limited $N_f$ locations (which corresponds to the observation points we need in the training of NP) in the domain. We are interested in learning the high-fidelity solution in the entire domain, for any instantiation of random variable $\mu$. 

We use 4 layers and 128 neurons per layer for both encoder and decoder in the neural network. The context points and target points are again evenly distributed in the domain $\Omega = [0,1]\times [0,1]]$. We choose 5, 10, and 20 context/target points in the training. The learning rate is $5e-4$. The batch size is $50$.  The CPU time in each training epoch is around 0.18 second.

The performances of the test sample are investigate using mean of relative $L^2$ errors
\begin{equation*}
	{  \frac{1}{N_s} \sum_{i=1}^{N_s}  \frac{|| u_{\text{pred, i}} - u_{H,i} ||_2 }{ ||u_{H,i} ||_2} }
\end{equation*}
and mean of relative weighted energy errors
\begin{equation*}
	{  \frac{1}{N_s} \sum_{i=1}^{N_s } \frac{\int_{\Omega} \kappa |\nabla (u_{\text{pred, i}} - u_{H,i}) |^2 }{\int_{\Omega} \kappa |\nabla u_{H,i} |^2} }. 
\end{equation*}
where $N_s$ is the number of samples, $u_{pred,i}$ is the mean predictions from the network for pressure, $u_{H,i}$ is true reference pressure solution.

The results are presented in Table \ref{tab:elliptic-f}. One can observe that with the physical constraints, both mean relative $L^2$ errors and energy errors can be reduced a lot in all three cases when taking different number context points. We remark that the number of degrees of freedom for the solution $u^H$ is $676$. Taking 20 observational points (approximately 3\% of all points) can give us pretty good predictions. 

\begin{figure}[!htp]
	\begin{center}
		\includegraphics[scale=0.37]{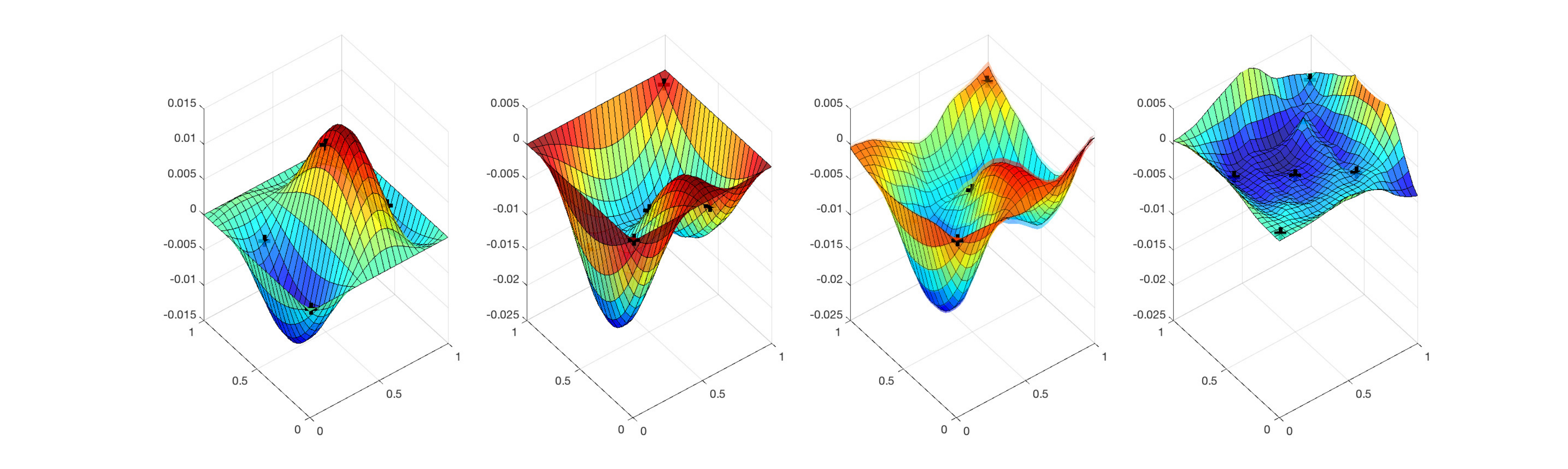}
		\includegraphics[scale=0.37]{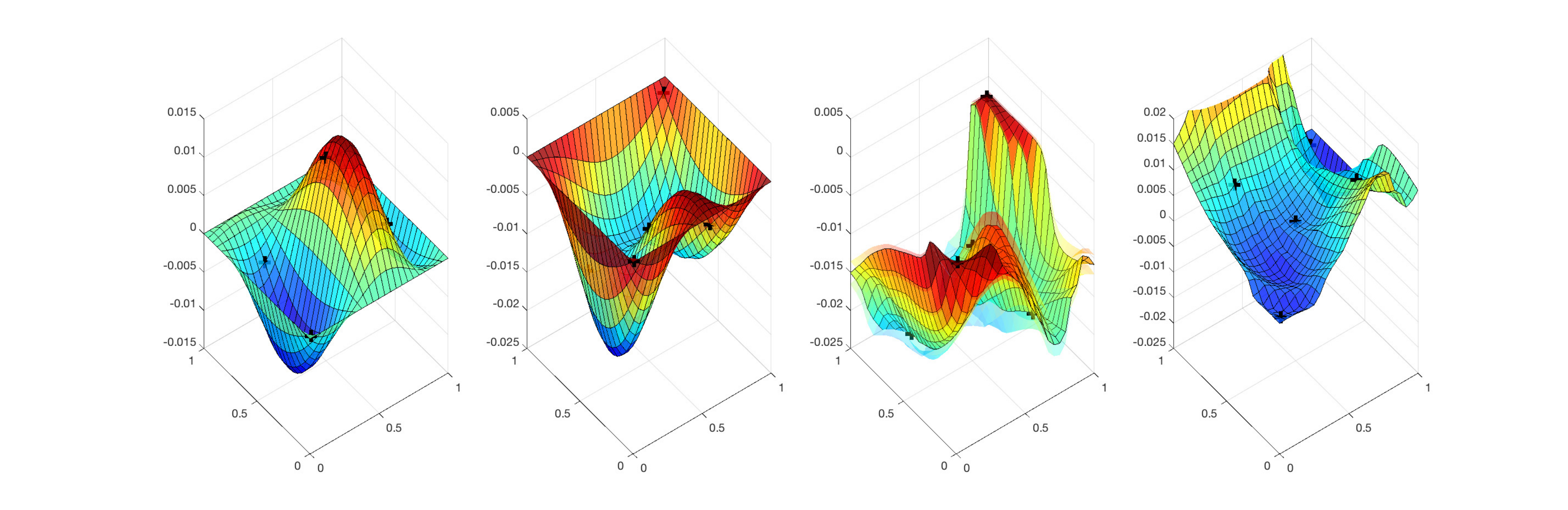}
	\end{center}
	\caption{The errors for a specific validation sample at the last epoch. Using 5 contexts points. Top row: with physical constraints. Bottom row: without physical constraints. From left to right: low fidelity solution, high fidelity solution, predicted solution, difference between high and predicted solution.} \label{fig:elliptic-f}
\end{figure}

An illustration of a specific sample prediction is depicted in Figure \ref{fig:elliptic-f}. In the figure, the first two columns refer to the low and high fidelity reference pressure, the third subplot from the left shows the mean of predictions. The uncertainties are presented by light blue shading in the third subplot, it doesn't show very clearly since the standard deviation actually is very small. The last columns are the differences between the mean prediction and reference solution. The top row refers to the results with physical constraints, the second row denotes the results without physical constraints. It shows that the predictions with physical constraints are much more accurate.

\begin{table}[ht!]
	\centering
	\begin{tabular}{|c|c|c|}
		\hline
		context/target points \# &$L^2$ (with PC)(\%)&  $L^2$ (without PC)(\%)  \\ \hline
		5 /5  &6.3  &13.53    \\ \hline
		10 /10 &2.83 &4.33   \\ \hline
		20 /20 &1.59  &3.75   \\ \hline
	\end{tabular}
	\begin{tabular}{|c|c|c|c|}
		\hline
		context/target points \#  &  energy (with PC)(\%)&  energy (without PC)(\%)  \\ \hline
		5 /5   & 7.67 & 14.45   \\ \hline
		10 /10 &3.64 &7.99    \\ \hline
		20 /20  &2.63  &4.53 \\ \hline
	\end{tabular}
	\caption{Relative mean errors comparison between true and predicted solutions.}\label{tab:elliptic-f}
\end{table}

%


\subsection{Elliptic PDE: inverse problem. }

\subsubsection{Identifying smooth random permeability field}

In practice, the permeability fields in the model are usually unknown. In this section, we will identify the permeability fields in the flow problem.  Consider the elliptic equation, 
\begin{equation} \label{eq:elliptic}
	-div (\kappa(x; \mu) \nabla u ) = f  \;\; \text{in} \;\; \Omega
\end{equation}
with Dirichlet boundary condition $u = 0$ on $\partial \Omega_1$ and free Neumann boundary condition $\frac{\partial u } {\partial n} = 0$ on $\partial \Omega_2$, where $\partial \Omega_1$ is the top and bottom boundaries, $\partial \Omega_2$ is the left and right boundaries on the square domain $\Omega$. Here $\kappa(x; \mathbf{\xi})$ is the spatial varied diffusion coefficient.

For the high-fidelity models, we choose the permeability field $\kappa(x; \mu)$ can be represented as follows:
\[
\kappa^H(x; \mu) = \kappa_0 +\displaystyle{\sum_{j=1}^{P_H}} \mu_j  \sqrt{\xi_j} \Phi_j(x)
\] 
where $ \kappa_0$ is a constant permeability, denotes the mean of the random field. $\displaystyle{\sum_{j=1}^{P_H} \mu_j \sqrt{\xi_j} \Phi_j(x)}$ denotes a random contribution obtained from Karhunen-Loeve expansion, and describes the uncertainty in the permeability field. $\mu_j $ are random numbers drawn from i.i.d $N(0,1)$. $(\sqrt{\xi_j}, \Phi_j(x))$ are the eigen-pairs obtained from a Gaussian covariance kernel:
\begin{equation*}
	\text{Cov} (x_i, y_i; x_j, y_j) = \sigma \exp(\frac{|x_i -x_j|^2}{l_x^2} -\frac{|y_i -y_j|^2}{l_y^2}  )
\end{equation*}
where we choose $[l_x, l_y]=[0.05, 0.05]$ and $\sigma=10$ in our example.

For the low permeability data, we just truncate $\kappa^L(x; \mu) =  \displaystyle{\sum_{j=1}^{P_L}} \mu_j \sqrt{\xi_j} $ at $P_L <P_H$. In the examples, we choose $P_L =2$, and $P_H=10$.

One can solve \eqref{eq:elliptic} using finite element method. In this case, one needs to reconstruct fine scale stiffness matrix for every $\kappa(x; \mu)$. Since this is an affine case, actually, we can compute $A_f(\kappa_0)$, and $A_f(\Phi_j(x))$ ($j = 1, \cdots, P_h$) once and then store them for future use. These fine scale stiffness matrices have size $N_f \times N_f$, where $N_f$ is the number of fine degrees of freedom. For an arbitrary $\kappa(x; \mu) = \kappa_0 +\displaystyle{\sum_{j=1}^{P_H}} \mu_j \sqrt{\xi_j} \Phi_j(x)$, we have
\[
A_f(\kappa(x; \mu)) = A_f(\kappa_0) +\displaystyle{\sum_{j=1}^{P_H}} \mu_j  A_f(\sqrt{\xi_j} \Phi_j(x))
\] 
and the corresponding fine scale solution is obtained by solving the large system
\[
A_f(\kappa(x;\mu)) u_f(\kappa(x; \mu)) = b.
\]

We aim to use the low-fidelity solutions and some limited observations from high-fidelity models to reconstruct the high-fidelity solutions as well as the permeability field. For training purposes, we still generate high-and low-fidelity solutions by solving the equations on a 26 by 26 grid, and observation data are the high-fidelity simulated solutions in selected locations. 

In this example, since the permeability fields are linear combinations of the weighted eigenfunctions from a given kernel, we can save the weighted eigenfunctions $ \sqrt{\xi_j} \Phi_j(x)$, and learn the random coefficients $\mu_j$ only. The desired permeability can then be reconstructed easily. We remark that, if the permeability fields are non-smooth and cannot be represented as linear combinations of these eigenfunctions, we will just learn them pixel-wisely.

\begin{figure*}[!htp]
	\centering
	\begin{subfigure}[t]{0.8\textwidth}
		\centering
		\includegraphics[scale=0.45]{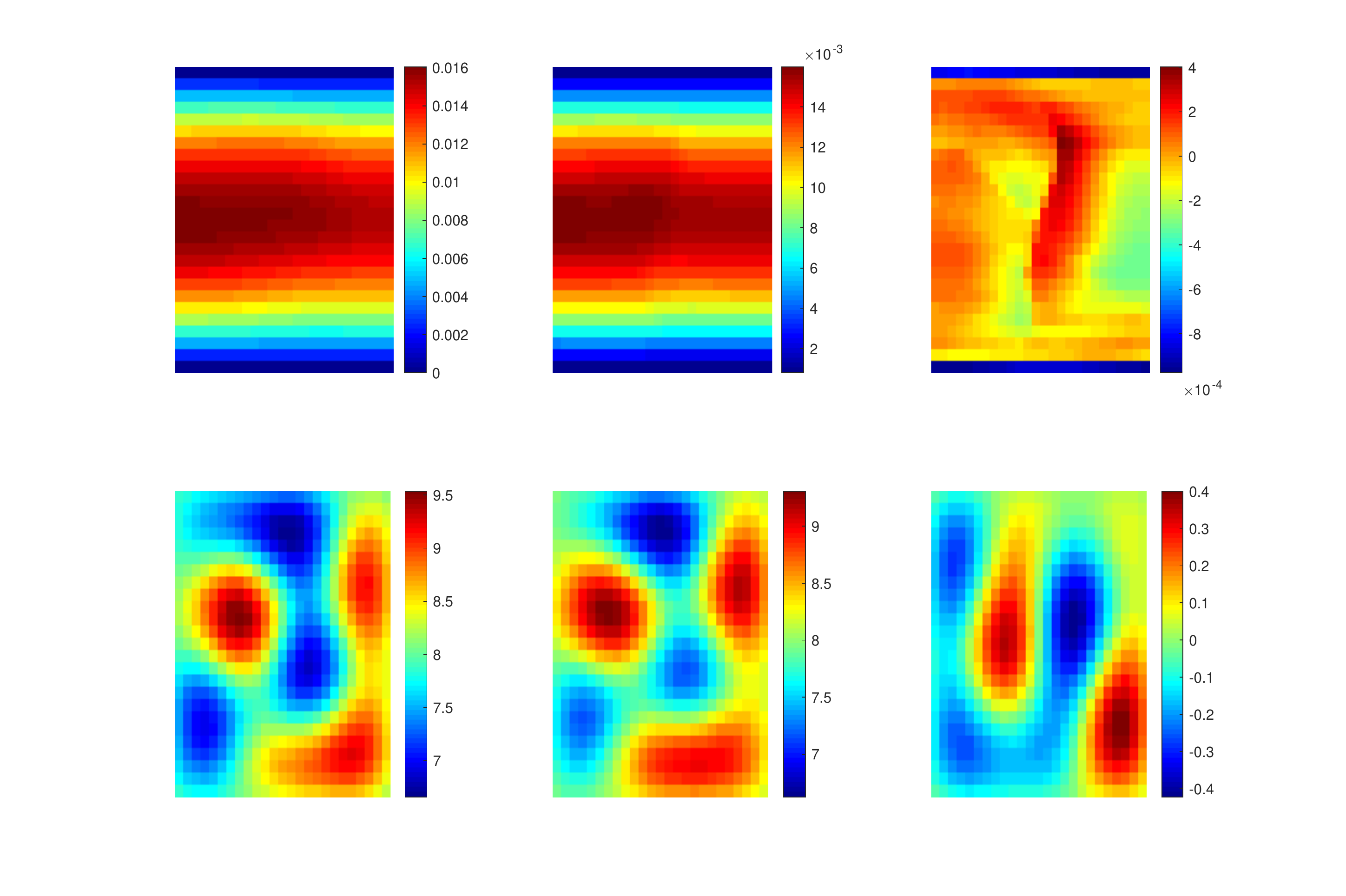}
		\caption{test sample 1}
	\end{subfigure}%
	
	
	
	\begin{subfigure}[t]{0.8\textwidth}
		\centering
		\includegraphics[scale=0.45]{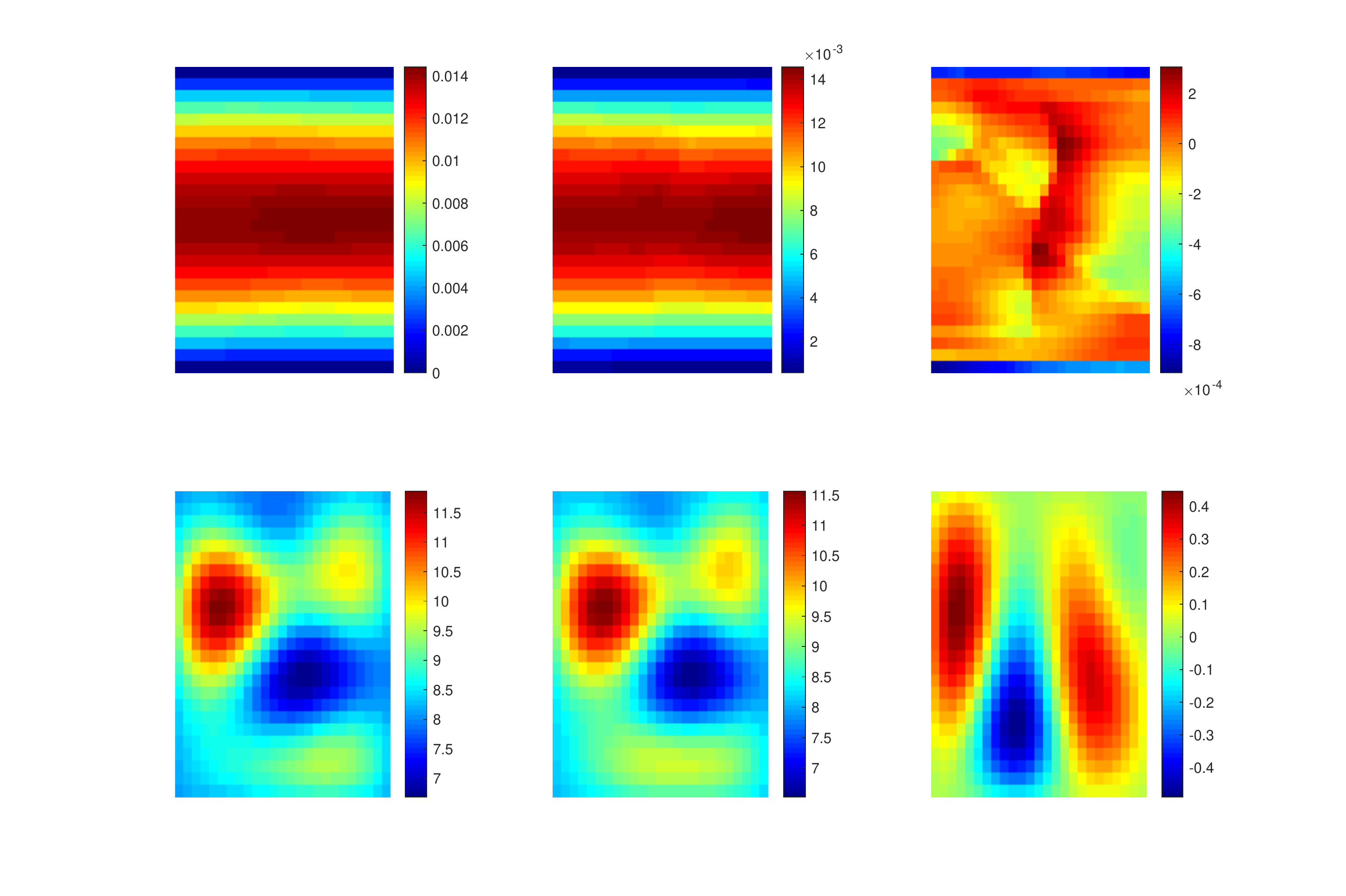}
		\caption{test sample 2}
	\end{subfigure}
	\caption{In each subfigure. Top row: high fidelity solution. Bottom row: permeability fields. Left column: reference. Middle column: predictions from our proposed methods. Right column: differences between the reference and prediction.}\label{fig:2d-inverse}
\end{figure*}

In Figure \ref{fig:2d-inverse}, we present the learning results for some samples when we use 15 context points. In each case, we obtain the high-fidelity solutions with uncertainty and the underlying high-fidelity permeability fields. Compared the reference high fidelity solutions/permeabilities with the predicted mean of high-fidelity solutions/permeabilities, we observe excellent matches.

\begin{table}[ht!]
	\centering
	\begin{tabular}{|c|c|c|c|}
		\hline
		context/target points \# & 5/5 & 10 /10 &15 /15
		\\ \hline
		solution relative $L^2$ error(\%)&3.65 &2.86 &1.76
		\\ \hline
		solution relative energy error(\%)&30.45 &15.07  &12.77
		\\ \hline
		relative permeability  $l_2$ error(\%) &12.73 & 12.59 &8.77  \\ \hline
	\end{tabular}
	\caption{Relative mean errors comparison between true and predicted solutions, as well as true and predicted permeability fields.}\label{tab:2d-inverse}
\end{table}

\subsubsection{Identifying channelized random permeability field}

In the last example, we would like to identify channelized random permeability field. The idea is the same as in the previous section. However, we will consider non-smooth  permeability fields, which cannot be represented as linear combinations of some random fields. We will learn the permeability pixel-wisely. In this example, we use zero Dirichlet boundary conditions. The right hand source term is now fixed to be a constant $f=1$ in the entire domain.

\begin{figure*}[!htp]
	\centering
	\includegraphics[scale=0.25]{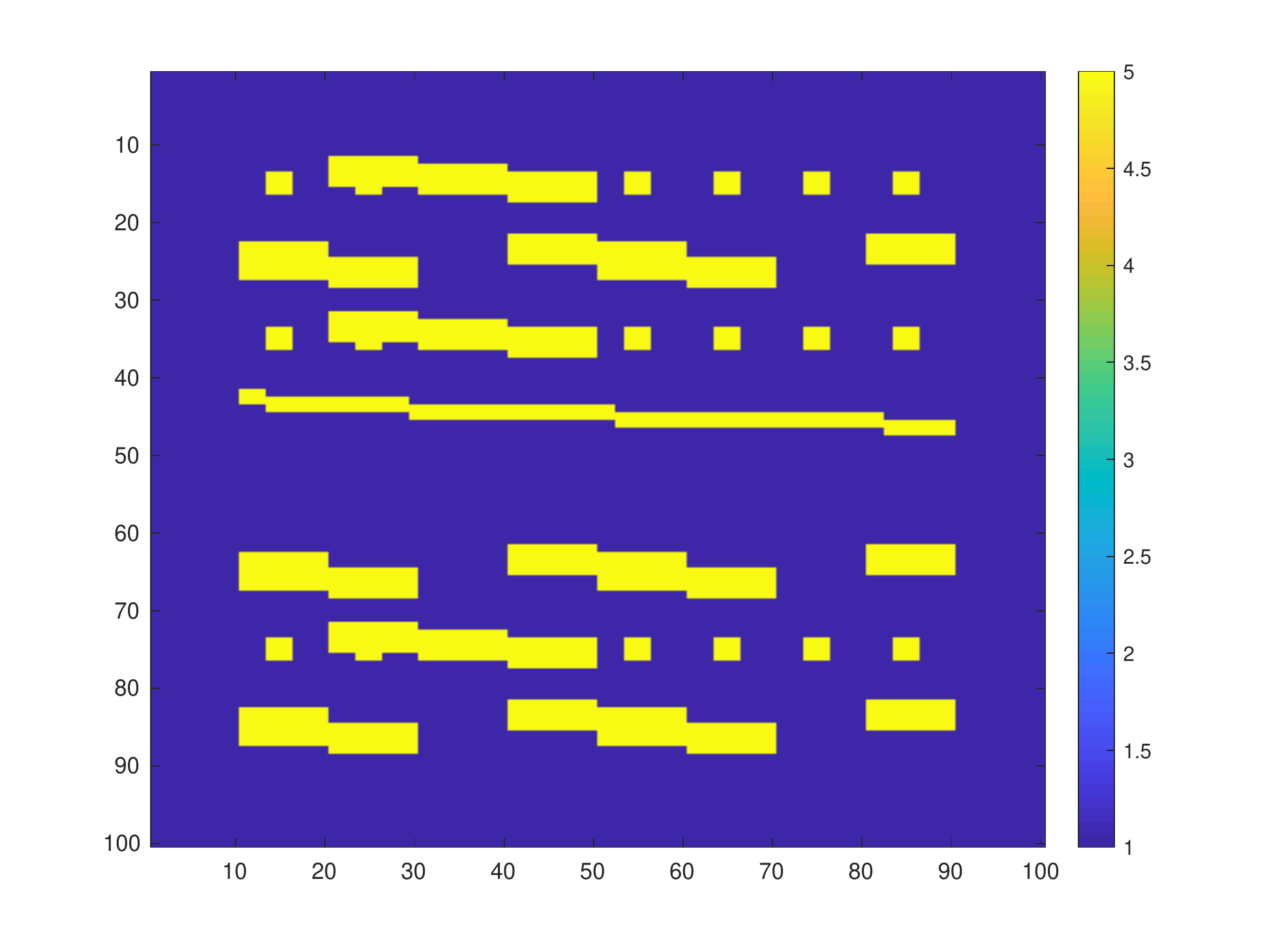}
	\caption{The channelized permeability field.}\label{fig:channel}
\end{figure*}

For the high fidelity models, we choose the permeability field $\kappa(x; \mu)$ to be:
\[
\kappa^H(x; \mu) = \kappa_{c} + \kappa_0 +\displaystyle{\sum_{j=1}^{P_h}} \mu_j  \sqrt{\xi_j} \Phi_j(x)
\] 
where $  \kappa_0 +\displaystyle{\sum_{j=1}^{P_H}} \mu_j  \sqrt{\xi_j} \Phi_j(x)$ is same as before, $P_H=10$. $\kappa_{c}$ is a channelized field as shown in Figure \ref{fig:channel}. The low fidelity permeability will only be $\kappa^L(x; \mu) =  \kappa_0 +\displaystyle{\sum_{j=1}^{P_H}} \mu_j  \sqrt{\xi_j} \Phi_j(x)$ without knowing the channels. Again, $\mu_j$ are randomly generated from i.i.d.

\begin{figure*}[!htp]
	\centering
	\begin{subfigure}[t]{0.8\textwidth}
		\centering
		\includegraphics[scale=0.45]{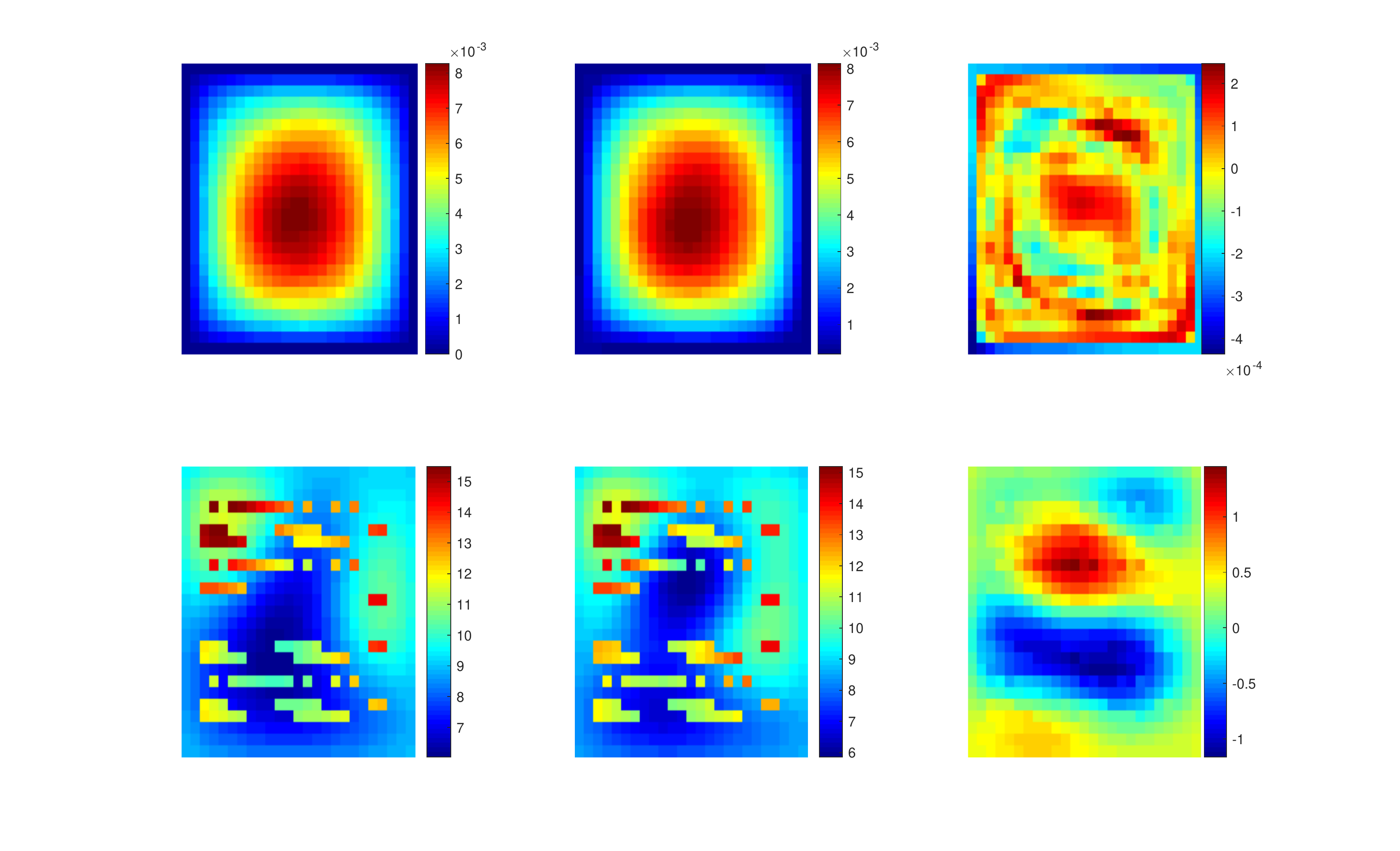}
		\caption{test sample 1}
	\end{subfigure}%
	
	\begin{subfigure}[t]{0.8\textwidth}
		\centering
		\includegraphics[scale=0.45]{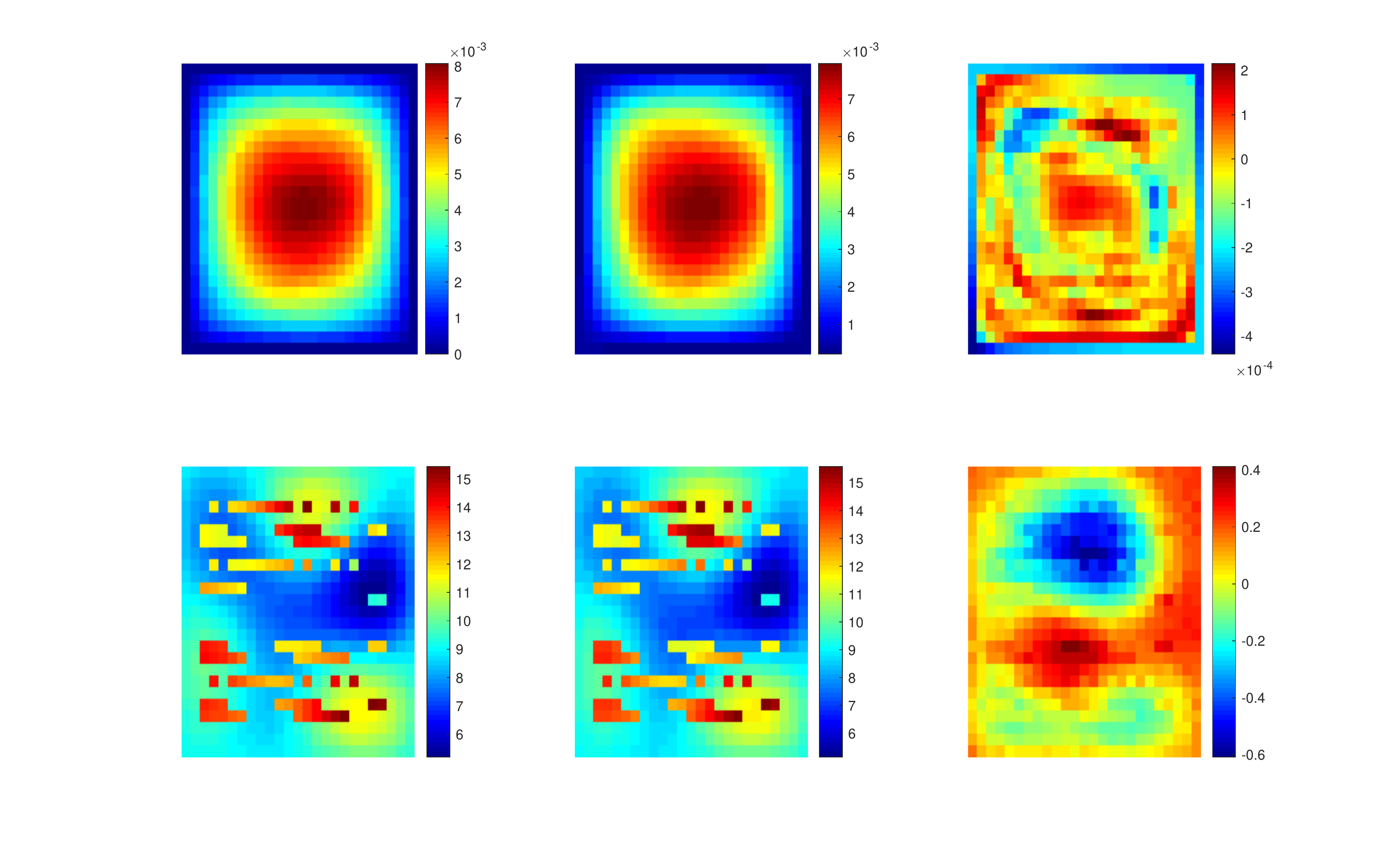}
		\caption{test sample 2}
	\end{subfigure}
	
	\caption{In each subfigure: Top row: high fidelity solution. Bottom row: permeability fields. Left column: reference. Middle column: predictions from our proposed methods. Right column: differences between the reference and prediction.}\label{fig:2d-inverse-channel}
\end{figure*}

In this network, we use additional 4 layers with 20 neurons per layer for the approximation of permeability in inverse process, please refer to the diagram in Figure \ref{fig:network}. 
Some test results are presented in Figure \ref{fig:2d-inverse-channel} and Table \ref{tab:2d-inverse-channel}. Even we now are learning the high permeability field pixel-wisely, we still notice very good matches between the reference high-fidelity solutions/permeability fields with predicted mean of high-fidelity solutions/permeability fields. However, since there are now more constraints in the loss function as shown in \eqref{eq:geco_loss_inverse}, it's harder to balance, especially$F^{H*} = L(\kappa^*, y^(H*))$ in the term $||F^H - F^{H*}||$ depends on both the predictions of $y^(H*)$ and $\kappa^*$. As we increase the number of context points, the errors in general decrease. We also remark that the physical constraints are necessary here in order to identify the unknown parameters in the PDE, so we don't have a comparison between the results with PC and without PC.

\begin{table}[!htp]
	\centering
	\begin{tabular}{|c|c|c|c|}
		\hline
		context/target points \#  & 10 /10 &15 /15 & 20/20
		\\ \hline
		solution relative $L^2$ error(\%)&1.9 &1.6 &1.7
		\\ \hline
		solution relative energy error(\%)&17.6 &15.0  & 12.3
		\\ \hline
		relative permeability  $l_2$ error(\%) &6.1 & 3.8 & 4.2 \\ \hline
	\end{tabular}
	\caption{Relative mean errors comparison between true and predicted solutions, as well as true and predicted permeability fields.}\label{tab:2d-inverse-channel}
\end{table}

\section{Conclusion}\label{sec:conclusion}

In this work, we propose MFPC-nets which join sufficient cheaply obtained low-fidelity data together with limited expensive high-fidelity data to train a network on top of the idea of neural process, and additionally impose physical laws of the underlying problem in loss functions. 
With a context set of input-output pairs, the proposed network learns to adapt their priors to the given data by minimizing the evidence lower bound constrained on physics. It can take multiple data sets and learn a distribution over functions. The loss function consisting of both parameters of the network and Lagrange multipliers corresponding to the constraints is optimized using stochastic gradient descent, and Lagrange multipliers are updated based on the constraints during training. After the learning process, one can obtain the correlation between low- and high-fidelity data for a class of models instead of a fixed one. Given a few low fidelity observations from a new model, the network is capable of rapid generalization to new observations and can estimate the uncertainty in their predictions. 

We test the performance of the proposed methods on several numerical examples.
Our first test is an one-dimensional pedagogical example, where the correlations between two fidelity models are nonlinear and stochastic. We observe that when there are only single fidelity data available, the network requires much more context points to get reasonable results. However, when the network takes multi-fidelity data as training samples, only with very few context points, it can provide very accurate and robust results.  Adding the physical constraints regarding the second derivative of the data can help to reduce the mean prediction errors significantly. Especially when the number of training samples or the number of data points is limited, adding physical constraints will produce much better results compared with the case without physical constraints. As we increase the number of context points, the constraints can be achieved faster. 

We then present examples regarding two-dimensional elliptic PDEs. We consider both forward and inverse problems. The forward problem learns the high-fidelity solutions from random source terms. The inverse problems learn the solutions as well as identify the smooth or channelized permeability fields of the high-fidelity model. The physical constraints are imposed using the weak form of the PDE. In the forward problem, the source terms are different and contain random parameters for the low/high fidelity model. The context points are the measurements/observations for both source terms and low/high fidelity solutions are chosen to be at the same locations, which are some evenly distributed points in the computational domain. The results show similar conclusions as in the one-dimensional case. More importantly, we adjust the architecture of the network and apply to inverse problems. For the second-order PDEs, the permeability coefficients usually contain uncertainly and are unknown in practice. We identify the permeability by adding additional constraints on the permeability and employ the mean predictions of the permeability further to realize the constrains on the physical equations in the loss. Both smooth and channelized permeability are considered, and the simulation results show good accuracy.

\section*{Acknowledgment}
We gratefully acknowledge the support from the National Science Foundation (DMS-1555072, DMS-1736364, CMMI-1634832 and CMMI-1560834), and Brookhaven National Laboratory Subcontract 382247, ARO/MURI grant W911NF-15-1-0562 and U.S. Department of Energy (DOE) Office of Science Advanced Scientific Computing Research program DE-SC0021142.

\bibliographystyle{siam} 
\bibliography{references}

\appendix
\section{Appendix} \label{sec:append_a}
In order to be label-efficient, we can also apply the active learning idea during the learning to choose batch samples. That is, we aim to find the most representative queries to be labeled in the training. Let $\Big( (x, y^L), y^H)\Big)$ be a sample pair belonging to the pool of labeled data $\Big((X, Y^L), Y^H)\Big)$. Suppose there exist a large pool of unlabeled samples $(x_P, y^L_P) \in (X_P, Y^L_P)$. We would like to query the most informative samples from these unlabeled pool based on some acquisition functions and  add them into the training in an iterative manner.

Let $\mathcal{NP}$ be the current neural network model, and $N_b$ be number of samples in a batch. The active learning idea is as follows:

\begin{algorithm}
	\caption{Active learning}\label{alg:AL}
	\begin{algorithmic}[1]
		\Procedure{AL}{ $\mathcal{NP}$, $\Big((X, Y^L), Y^H)\Big)$,  $X_P, Y^L_P$, $N_b$}
		\While{$ (X_P, Y^L_P) \neq \emptyset$}
		\State Train $\mathcal{NP}$ on current labeled data $\Big((X, Y^L), Y^H)\Big)$
		\State Apply $\mathcal{NP}$ on unlabeled data $X_P, Y^L_P$
		\State Rank the predicted variance of $ Y^H_P$
		\State Choose the largest $N_b$ samples $(x_i, y^L_i)_{i=1,\cdots N_b} \subset (X_P, Y^L_P)$ 
		\State Acquire $y^H_i$, $i=1,\cdots N_b$.
		\State $(X_P, Y^L_P) \gets (X_P, Y^L_P)  \ (x_i, y^L_i)_{i=1,\cdots N_b}$
		\State $(X, Y^L) \gets (X, Y^L)  \cup  (x_i, y^L_i)_{i=1,\cdots N_b}$
		
		\EndWhile
		\EndProcedure
	\end{algorithmic}
\end{algorithm}

One can also randomly add data from unlabeled sets to the training data, that is, steps $5-6$ in Algorithm \ref{alg:AL} can be replaced by a random selection function.

\end{document}